\documentclass[useAMS,usegraphicx,usenatbib]{mn2e} 
\usepackage{longtable}
\usepackage{natbib}
\usepackage{amsmath}
\usepackage{color}
\usepackage{url}
\usepackage{ulem}
\usepackage{multirow}
\usepackage{amsmath}
\usepackage{wasysym}
\usepackage{amssymb}
\usepackage{pifont}

\bibpunct{(}{)}{;}{a}{}{,}

\def\apjl{ApJ}%
\def\apjs{ApJS}%
\def\ao{Appl.~Opt.}%
\def\aap{A\&A}%
\title[BASS XII: Coronal properties of AGN]{BAT AGN Spectroscopic Survey -- XII.  
The relation between coronal properties of Active Galactic Nuclei and the Eddington ratio}
\author[C. Ricci et al.]{C. Ricci$^{1,2,3}$\thanks{E-mail:
claudio.ricci@mail.udp.cl} , L. C. Ho$^{3,4}$, A. C. Fabian$^{5}$, B. Trakhtenbrot$^{6}$, M. J. Koss$^{7}$,\newauthor
Y. Ueda$^{8}$, A. Lohfink$^{9}$, T. Shimizu$^{10}$, F. E. Bauer$^{11,12,13}$, R. Mushotzky$^{14}$,\newauthor K. Schawinski$^{15}$, S. Paltani$^{16}$, I. Lamperti$^{17}$, E. Treister$^{11}$,  K. Oh$^{8}$ \\ 
$^{1}$N\'ucleo de Astronom\'ia de la Facultad de Ingenier\'ia, Universidad Diego Portales, Av. Ej\'ercito Libertador 441, Santiago, Chile\\
$^{2}$Kavli Institute for Astronomy and Astrophysics, Peking University, Beijing 100871, China\\
$^{3}$Chinese Academy of Sciences South America Center for Astronomy, Camino El Observatorio 1515, Las Condes, Santiago, Chile\\
$^{4}$Department of Astronomy, School of Physics, Peking University, Beijing 100871, China\\
$^{5}$Institute of Astronomy, Madingley Road, Cambridge CB3 0HA, UK\\
$^{6}$Department of Physics, ETH Zurich, Wolfgang-Pauli-Str. 27, CH-8093 Zurich, Switzerland\\
$^{7}$Eureka Scientific Inc., 2452 Delmer St. Suite 100, Oakland, CA 94602, USA\\
$^{8}$Department of Astronomy, Kyoto University, Kyoto 606-8502, Japan\\
$^{9}$Department of Physics, Montana State University, Bozeman, MT 59717-3840, USA\\
$^{10}$Max-Planck-Institut f\"{u}r extraterrestrische Physik, Postfach 1312, 85741, Garching, Germany\\
$^{11}$Instituto de Astrof\'{\i}sica, Facultad de F\'{i}sica, Pontificia Universidad Cat\'{o}lica de Chile, Casilla 306, Santiago 22, Chile\\
$^{12}$Space Science Institute, 4750 Walnut Street, Suite 205, Boulder, Colorado 80301, USA\\
$^{13}$Millenium Institute of Astrophysics, Santiago, Chile\\
$^{14}$Department of Astronomy and Joint Space-Science Institute, University of Maryland, College Park, MD 20742, USA\\
$^{15}$Institute for Particle Physics and Astrophysics, ETH Zurich, Wolfgang-Pauli-Str. 27, CH-8093 Zurich, Switzerland\\
$^{16}$Department of Astronomy, University of Geneva, ch. d'Ecogia 16, CH-1290 Versoix, Switzerland\\
$^{17}$Astrophysics Group, Department of Physics and Astronomy, University College London, 132 Hampstead Road, London NW1 2PS, UK\\
 }
\begin{document}
\date{Received; accepted}

\pagerange{\pageref{firstpage}--\pageref{lastpage}} \pubyear{2017}

\maketitle

\label{firstpage}

\begin{abstract}
The bulk of the X-ray emission in Active Galactic Nuclei (AGN) is produced very close to the accreting supermassive black hole (SMBH), in a corona of hot electrons which up scatters optical and ultraviolet photons from the accretion flow. The cutoff energy ($E_{\rm C}$) of the primary X-ray continuum emission carries important information on the physical characteristics of the X-ray emitting plasma, but little is currently known about its potential relation with the properties of accreting SMBHs. Using the largest broad-band (0.3--150\,keV) X-ray spectroscopic study available to date, we investigate how the corona is related to the AGN luminosity, black hole mass and Eddington ratio ($\lambda_{\rm Edd}$). Assuming a slab corona the median values of the temperature and optical depth of the Comptonizing plasma are $kT_{\rm e}=105 \pm 18$\,keV and $\tau=0.25\pm0.06$, respectively. When we properly account for the large number of $E_{\rm C}$ lower limits, we find a statistically significant dependence of the cutoff energy on the Eddington ratio. In particular, objects with $ \lambda_{\rm Edd}>0.1$ have a significantly lower median cutoff energy ($E_{\rm C}=160\pm41$\,keV) than those with $\lambda_{\rm Edd}\leq 0.1$ ($E_{\rm C}=370\pm51$\,keV). This is consistent with the idea that radiatively compact coronae are also cooler, because they tend to avoid the region in the temperature-compactness parameter space where runaway pair production would dominate. We show that this behaviour could also straightforwardly explain the suggested positive correlation between the photon index ($\Gamma$) and the Eddington ratio, being able to reproduce the observed slope of the $\Gamma-\lambda_{\rm Edd}$ trend.

\end{abstract}	
               


\section{Introduction}\label{sect:introduction}
Accreting supermassive black holes (SMBHs) are known to ubiquitously produce radiation in the X-ray band. The X-ray emission of these Active Galactic Nuclei (AGN) is thought to be produced in a {\it corona}  of hot electrons, which up-scatters optical and UV photons into the X-ray band through inverse Compton scattering (e.g., \citealp{Haardt:1991qr,Haardt:1993cv,Merloni:2001qy,Merloni:2003kx,Liu:2015vn,Liu:2017qy}).  
The size of the X-ray corona has been shown to be relatively small ($5-10\,R_{\rm g}$, where $R_{\rm g}=GM_{\rm BH}/c^2$ is the gravitational radius for a SMBH of mass $M_{\rm BH}$) from the rapid X-ray variability (e.g., \citealp{McHardy:2005kc}), and the short timescales of X-ray eclipses (e.g., \citealp{Risaliti:2005kl,Risaliti:2011yo}). This has been also confirmed by microlensing studies (e.g., \citealp{Chartas:2009sy}), which have found a half-light radius of the corona of $\sim6\,R_{\rm g}$. Reverberation studies of X-ray radiation reprocessed by the accretion disk have suggested that the X-ray source is located very close to the SMBH and the accretion disk (e.g., \citealp{Fabian:2009hi,Zoghbi:2012jk,De-Marco:2013fx,Kara:2013wu,Reis:2013kq}), typically within $3-10\,R_{\rm g}$. Despite these advances in localization and size estimates of the X-ray source, its physical characteristics are still debated. Besides providing critical insights on the physics of the innermost regions of SMBHs, a clear understanding of the typical characteristics of the X-ray emitting plasma for different intervals of the accretion rate is extremely important to assess the impact of radiative heating \citep{Xie:2017fb} in the feedback process linking AGN to their host galaxies (e.g. \citealp{Ferrarese:2000kq,Gebhardt:2000fj,Schawinski:2006kq,Fabian:2012eq,Kormendy:2013uf,King:2015ys}).

X-ray spectroscopy, and in particular the study of the primary X-ray emission produced in the Comptonizing plasma, can provide important insights on the physical parameters of the corona, such as its temperature ($kT_{\rm e}$) and optical depth ($\tau$). The two main spectral parameters carrying information on the physical properties of the X-ray corona are the photon index ($\Gamma$) and the energy of the cutoff ($E_{\rm C}$; e.g., \citealp{Mushotzky:1993bf}). While the photon index has been routinely studied over the past two decades by observations carried out in the 0.3--10\,keV band, the cutoff energy has been more difficult to constrain, since it requires good-quality data above 10\,keV.  Indirect constraints on the cutoff energy have been obtained by \citet{Gilli:2007yg} who, studying the cosmic X-ray background (CXB, see also \citealp{Treister:2009qa}), showed that the mean cutoff energy should lie below 300\,keV; \citet{Treister:2005zr} and \citet{Ueda:2014ix} were able to reproduce the CXB assuming $E_{\rm C}=300\rm\,keV$; fitting the X-ray luminosity function of local AGN in four energy bands, \cite{Ballantyne:2014dq} found that the typical cutoff energy should be $E_{\rm C}\sim 200-450$\,keV. Spectroscopic studies carried out using the Gamma Ray Observatory/OSSE (e.g., \citealp{Zdziarski:1996qp,Johnson:1997hw}), BeppoSAX (e.g., \citealp{Nicastro:2000il,Dadina:2007sj}), {\it INTEGRAL} IBIS/ISGRI (e.g., \citealp{Beckmann:2009fk,Molina:2009vz,Lubinski:2010rb,Lubinski:2016ao,Ricci:2011yw,Panessa:2011pv,de-Rosa:2012pd,Malizia:2014zt}), {\it Swift}/BAT (e.g., \citealp{Vasudevan:2013wb}) and {\it Suzaku}/PIN (e.g., \citealp{Tazaki:2011xi}) were able to constrain the cutoff energies of several local bright AGN. 

More recently, in \cite{Ricci:2017bf}, we carried out the largest study of broad-band X-ray spectra (0.3--150\,keV) to date (836 AGN), showing that, in the large majority ($\simeq 80\%$) of the non-blazar AGN, the spectral slope of the 14--195\,keV emission is steeper than that in the 0.3--10\,keV band. This suggests that a high-energy cutoff is almost ubiquitous in AGN. The detailed broad-band X-ray spectral analysis of all sources of the sample showed that the median value of the cutoff energy of local AGN is $200 \pm 29$\,keV \citep{Ricci:2017bf}.
The recent launch of {\it NuSTAR} \citep{Harrison:2013lq} has greatly improved our understanding of cutoff energies, allowing to accurately constrain this parameter for a growing number of AGN, most of which reside at low redshifts (e.g., \citealp{Ballantyne:2014kc,Brenneman:2014mi,Matt:2014fv,Marinucci:2014fu,Balokovic:2015mi,Matt:2015fe,Parker:2014zp,Ursini:2015dk,Lohfink:2015ec,Lohfink:2017bq,Lanzuisi:2016pl,Kara:2017lq,Xu:2017kq,Tortosa:2017kq,Tortosa:2018lk,Tortosa:2018rm}). Exploiting the revolutionary capabilities of {\it NuSTAR}, \citeauthor{Fabian:2015db} (\citeyear{Fabian:2015db}; see also \citealp{Fabian:2017dq}) have shown that coronae lie close to the boundary of the region in the temperature--compactness parameter space which is forbidden due to runaway pair production (see \S\ref{sec:thetalplane}). 
Studying 19 {\it Swift}/BAT AGN with {\it NuSTAR}, \citet{Tortosa:2018rm} found no evidence of a significant correlation between $E_{\rm C}$ and black hole mass or Eddington ratio. However, the sample of bright AGN that are observed by {\it NuSTAR} and have reliable determinations of these key SMBH properties is still small, and does not allow to exclude the existence of relations between the coronal properties and the physical characteristics of the SMBH.

In order to improve our understanding of the properties of accreting SMBHs in the local Universe, our group has been systematically studying the properties of {\it Swift}/BAT AGN across the electromagnetic spectrum, in the framework of the BAT AGN Spectroscopic Survey (BASS\footnote{http://www.bass-survey.com}, \citealp{Koss:2017fp,Ricci:2017bf}). Previous publications based on BASS have studied the optical lines \citep{Berney:2015uq,Oh:2017zl}, the near-infrared emission \citep{Lamperti:2017kq}, the X-ray photon index \citep{Trakhtenbrot:2017bh} and the absorption properties \citep{Ricci:2015tg,Ricci:2017kl,Shimizu:2018fj} of {\it Swift}/BAT AGN.
Exploiting the rich multi-wavelength database available for BASS AGN, here we investigate the relation between the high-energy cutoff and the fundamental properties of AGN, such as their luminosity ($L$), black hole mass ($M_{\rm BH}$) and Eddington ratio ($\lambda_{\rm Edd}=L/L_{\rm Edd}$, see Eq.\,\ref{eq:eddratio}). 
The paper is structured as follows. In \S\ref{sect:sampledata} we introduce the sample used for this work, while in \S\ref{sec:Ecvsaccretion} we study how the cutoff energy is related to luminosity, black hole mass and Eddington ratio, showing that AGN accreting at high Eddington ratios ($\log \lambda_{\rm Edd} \geq -1$) typically have lower cutoff energies than those accreting at lower Eddington ratios ($\log \lambda_{\rm Edd} < -1$). In \S\ref{sec:thetalplane} we discuss how our sources are distributed in the temperature--compactness parameter space, and how this relates to the dissimilar typical cutoff energies of AGN populations accreting at different $\lambda_{\rm Edd}$. In \S\ref{sect:tau} we investigate the relation between the optical depth of the Comptonizing plasma and the properties of the accreting SMBH. In \S\ref{sec:GammaEddratiocorrelation} we show how the fact that AGN avoid the region in the temperature--compactness parameter space where runaway pair production takes place would produce the observed correlation between the photon index and the Eddington ratio. Finally, in \S\ref{sect:summary} we present our conclusions and summarise our findings.


\section{BASS: Sample and Data}\label{sect:sampledata}

The Burst Alert Telescope (BAT, \citealp{Barthelmy:2005uq}) on board the {\it Neil Gehrels Swift Observatory} \citep{Gehrels:2004kx} has been scanning the whole sky in the 14-195\,keV band since its launch in 2005, detecting 838 AGN in the first 70-months of operations \citep{Baumgartner:2013ee,Ricci:2017bf}. The multi-wavelength survey BASS has collected data in the radio, infrared, optical and X-rays for the large majority of these objects. In the following, we report on the X-ray (\S\ref{sect:xraydata}) and optical (\S\ref{sect:opticaldata}) data used for this work.

\subsection{X-ray data}\label{sect:xraydata}
The cutoff energies and the AGN luminosities used here are taken from the BASS X-ray catalogue \citep{Ricci:2017bf}, which reports the broad-band X-ray spectral properties for the 836 AGN detected by {\it Swift}/BAT in its first 70 months of operations ($\simeq 99.8\%$ of the total sample) for which soft X-ray (0.3--10\,keV) spectra were available. 
This was done by combining the 70-month averaged {\it Swift}/BAT spectra with shorter pointed observations carried out by {\it Swift}/XRT, {\it XMM-Newton}/EPIC, {\it Chandra}/ACIS, {\it Suzaku}/XIS and {\it ASCA} GIS/SIS. The spectral analysis was carried out over the entire 0.3--150\,keV range, using a total of 26 different spectral models, which include various emission components. The broad-band X-ray coverage allowed to recover several important properties of these AGN, such as their intrinsic X-ray luminosity, column densities, photon indices and cutoff energies. For further details on the spectral analysis we refer the reader to \citet{Ricci:2017bf}. We focus here only on the 317 unobscured [i.e., $\log(N_{\rm H}/{\rm cm}^{-2}) < 22$] AGN for which $E_{\rm C}$ could be constrained\footnote{For the remaining unobscured AGN the cutoff energy could not be constrained by the fit.} (228 lower limits and 89 values), to avoid possible degeneracies due to the additional spectral curvature introduced by heavy obscuration above 10\,keV.

\begin{figure}
\centering
\includegraphics[width=0.48\textwidth]{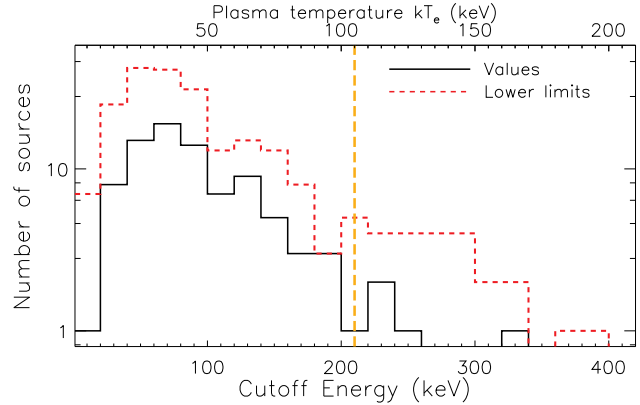}
  \caption{Histogram of the cutoff energy of unobscured ($N_{\rm H}<10^{22}\rm cm^{-2}$) sources from the {\it Swift}/BAT AGN catalog of \citet{Ricci:2017bf}. The temperature of the Comptonizing plasma was calculated assuming $kT_{\rm e}=E_{\rm C}/2$ (see \S\ref{sec:thetalplane}). The continuous black and dashed red lines illustrate the values and the lower limits, respectively. The vertical dashed orange line shows the median cutoff energy and plasma temperature of the sample ($E_{\rm C}=210 \pm 36$\,keV, i.e. $kT_{\rm e}=105\pm18$\,keV), calculated taking into account the lower limits.}
\label{fig:Ecut_unobs}
\end{figure}

\begin{figure*}
\centering
\includegraphics[width=0.48\textwidth]{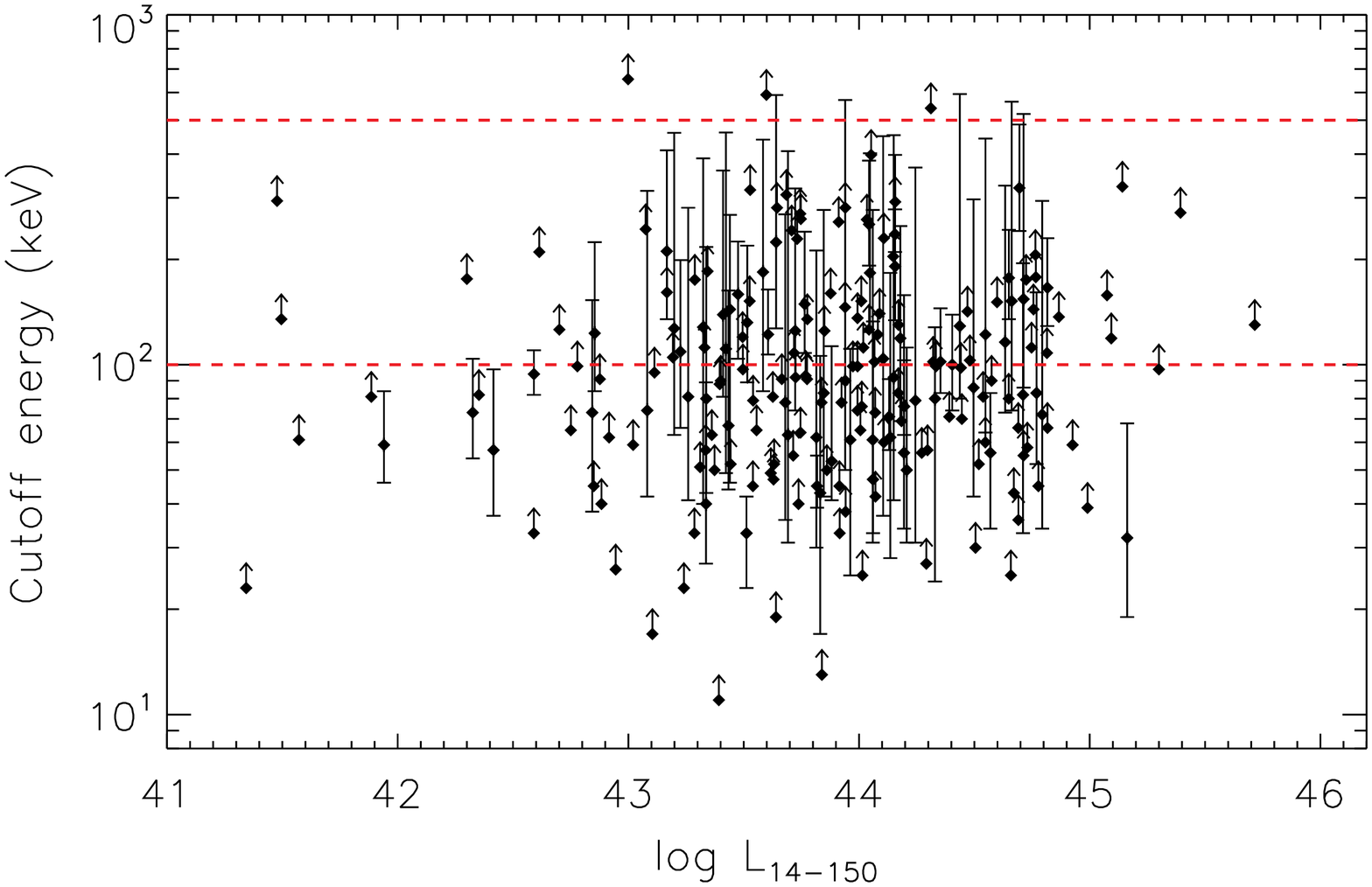}
\includegraphics[width=0.48\textwidth]{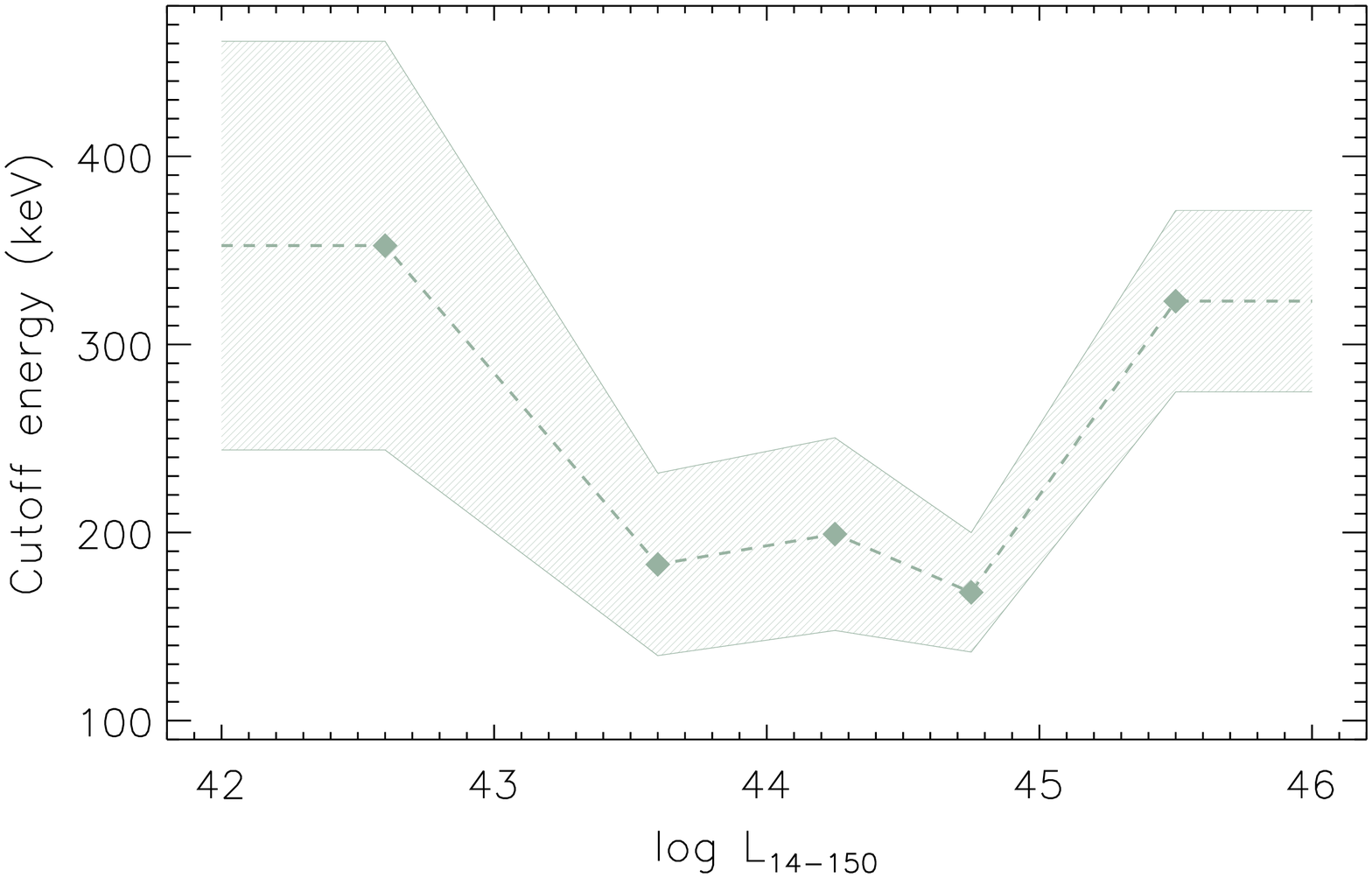}
  \caption{{\it Left panel:} Cutoff energy versus the 14--150\,keV intrinsic luminosity (in $\rm erg\,s^{-1}$) for the sources in our sample. The red dashed lines show the interval of cutoff energies shown in the right panel. {\it Right panel:} Median values of the cutoff energy for different bins of $L_{\rm 14-150}$, calculated including the lower limits using the Kaplan-Meier estimator within the \textsc{asurv} package. The shaded area corresponds to the median absolute deviation.}
\label{fig:Ecut_vsL}
\end{figure*}

\begin{figure*}
\centering
\includegraphics[width=0.48\textwidth]{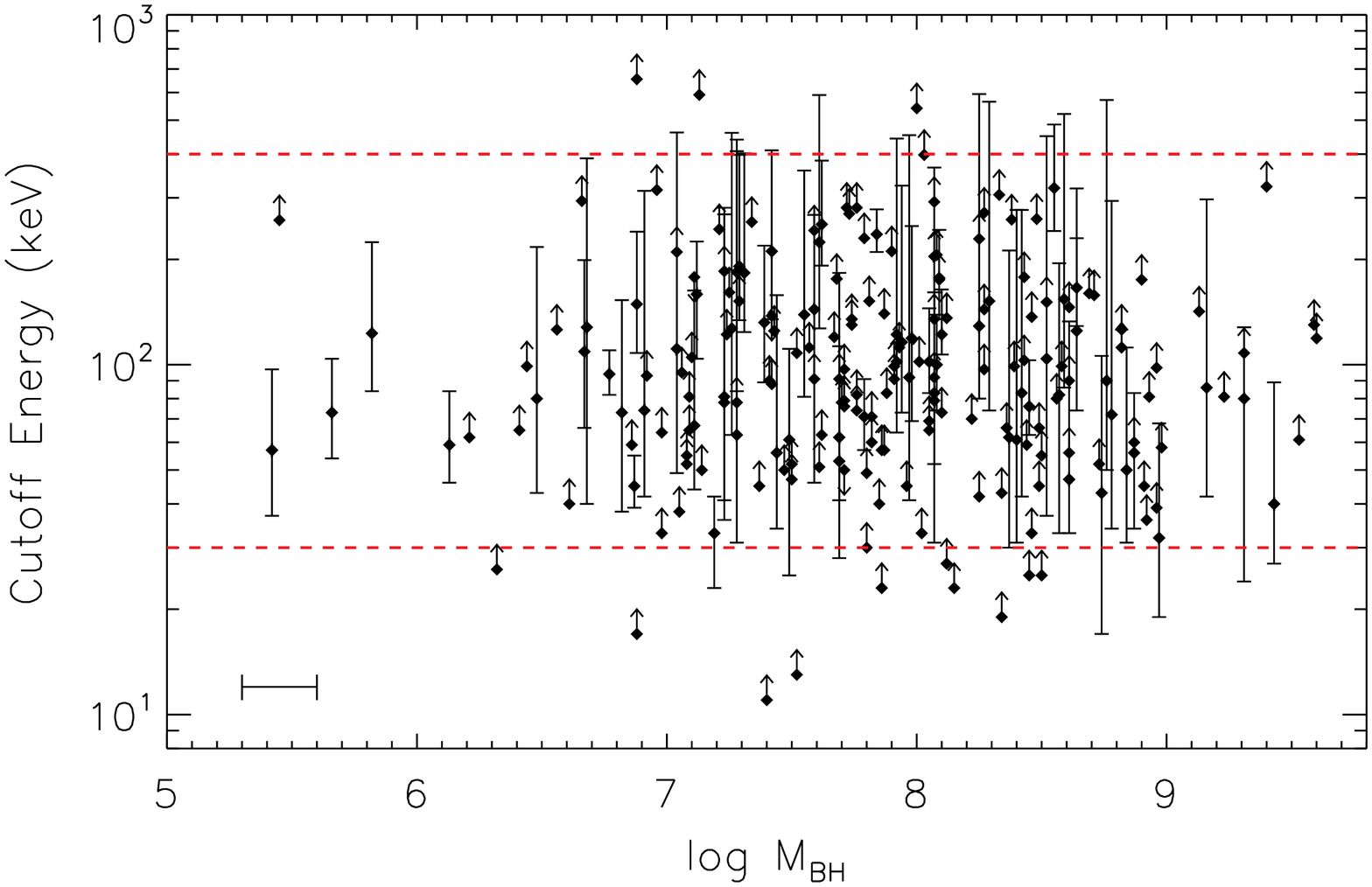}
\includegraphics[width=0.48\textwidth]{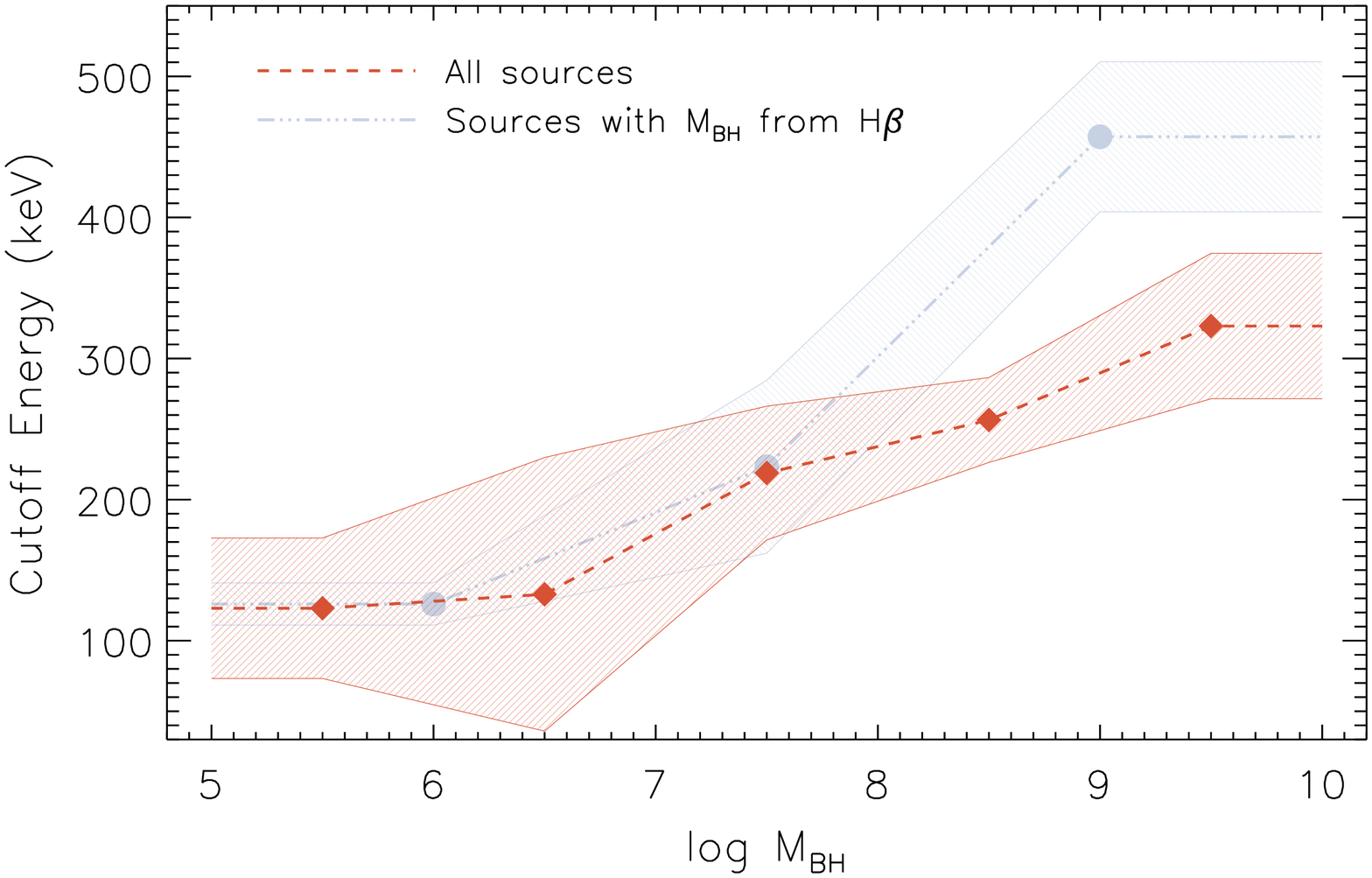}
  \caption{{\it Left panel:} Cutoff energy versus $M_{\rm BH}$ (in $M_{\odot}$) for the sources in our sample. The red dashed lines show the interval of cutoff energies shown in the right panel. The bar in the bottom left corner shows the typical uncertainty of $M_{\rm BH}$. {\it Right panel:} Median values of the cutoff energy for different intervals of $M_{\rm BH}$ for the whole sample (red dashed line) and for the objects for which $M_{\rm BH}$ was estimated using H$\beta$ (blue dot-dot dashed line). The medians were calculated including the lower limits using the Kaplan-Meier estimator within the \textsc{asurv} package. The shaded area represents to the median absolute deviation.}
\label{fig:Ecut_vsMBH}
\end{figure*}

\begin{figure*}
\centering
\includegraphics[width=0.48\textwidth]{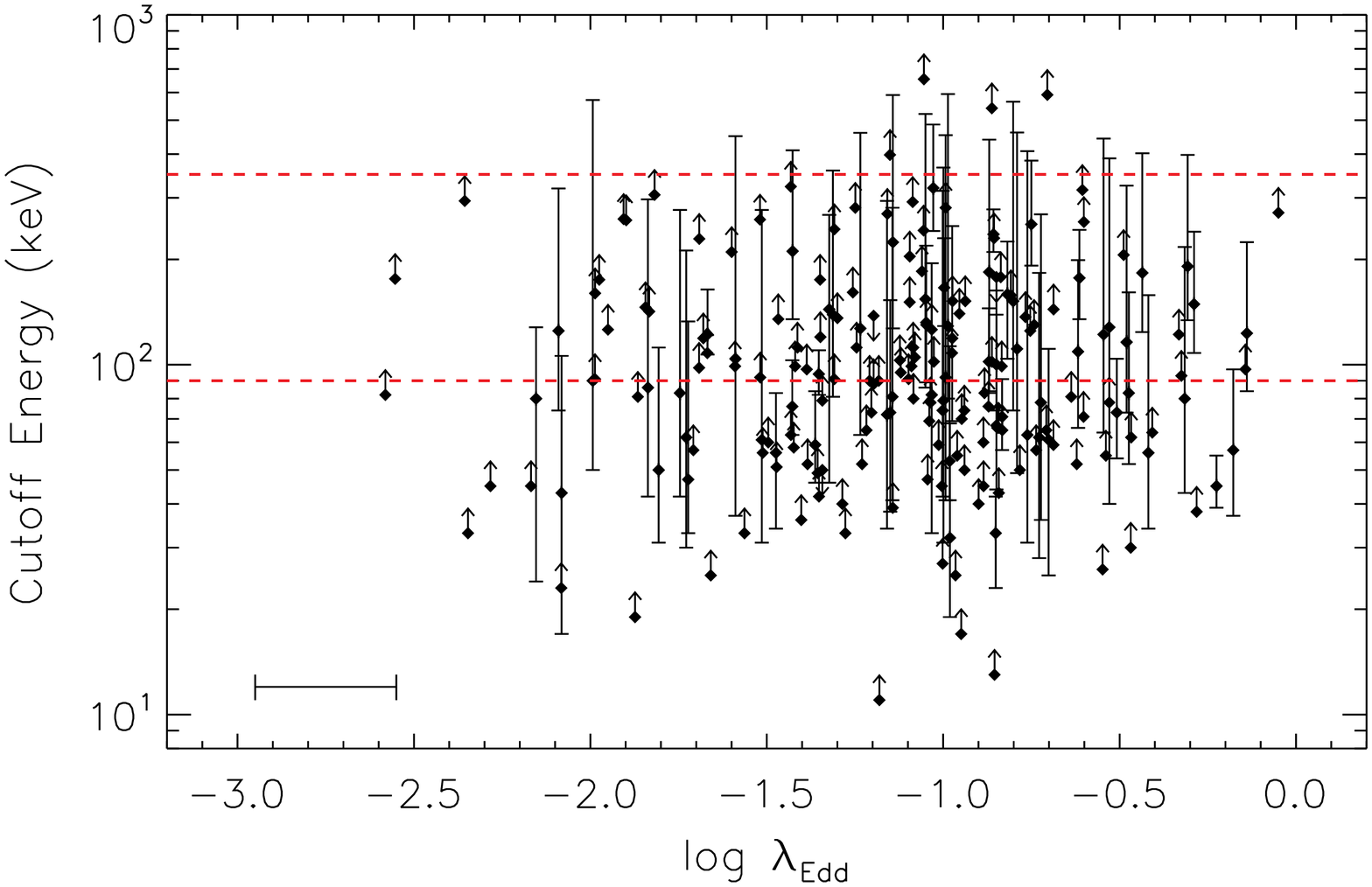}
\includegraphics[width=0.48\textwidth]{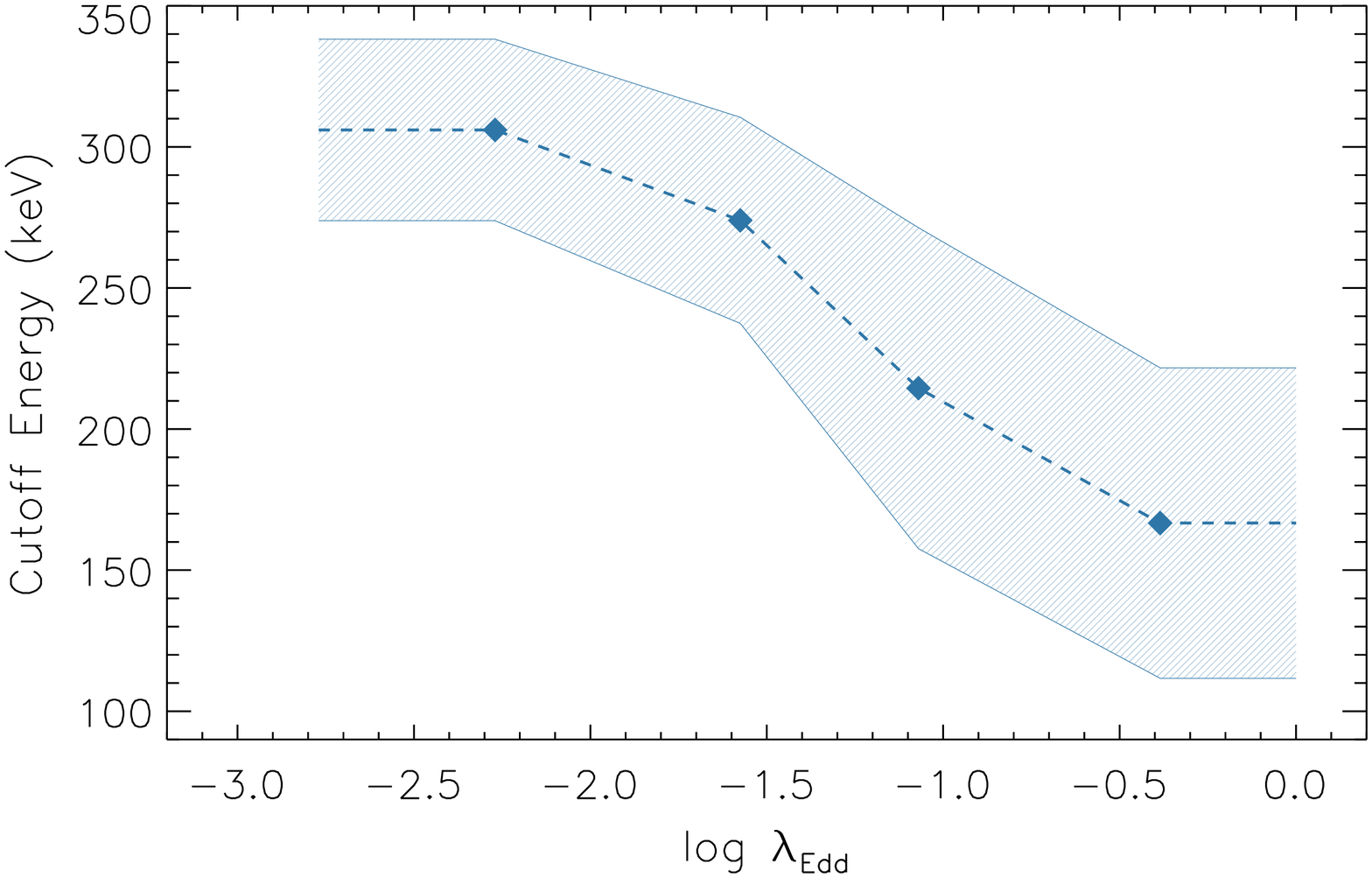}
  \caption{{\it Left panel:} Cutoff energy versus the Eddington ratio for the sources in our sample. The red dashed lines show the interval of cutoff energies shown in the right panel. The bar in the bottom left corner shows the typical uncertainty of $\lambda_{\rm Edd}$. {\it Right panel:} Median values of the cutoff energy for different intervals of $\lambda_{\rm Edd}$. The shaded area corresponds to the median absolute deviation.}
\label{fig:Ecut_vsEddratio}
\end{figure*}

\subsection{Optical data, black hole masses and Eddington ratio}\label{sect:opticaldata}
The analysis of the optical spectra of 642 {\it Swift}/BAT accreting SMBHs is reported in \cite{Koss:2017fp}, and allowed us to obtain black hole masses for 429 non-blazar AGN. Of these, 232 are unobscured, while 197 are obscured. The black hole masses were obtained through several fundamentally different approaches: i) ``direct'' methods (i.e., maser emission, spatially resolved gas- or stellar-kinematics, reverberation mapping); ii) single-epoch spectra of broad H$\beta$ and H$\alpha$ emission lines (e.g., \citealp{Trakhtenbrot:2012hq,Greene:2005wf}, respectively); iii) stellar velocity dispersions ($\sigma_{*}$) and the $M_{\rm BH}-\sigma_{*}$ relation \citep{Kormendy:2013uf}. 
Of the 232 unobscured AGN with black hole mass estimates, our broad-band X-ray spectral analysis could constrain cutoff energies for a total of 211 sources, of which 144 are lower limits and 67 are values. For these objects $M_{\rm BH}$ was obtained using broad H$\beta$ (144), reverberation mapping (31), broad H$\alpha$ (18), velocity dispersion (16), stellar (1) and gas (1) kinematics. These 211 AGN are a representative subset of sources of the BAT sample of unobscured AGN, having a very similar luminosity distribution. In the following we will use this as our final sample.

The Eddington luminosity was calculated using the following relation:
\begin{equation}\label{eq:eddratio}
L_{\rm Edd}=\frac{4\pi G M_{\rm BH} m_{\rm p}c}{\sigma_{\rm T}},
\end{equation}
where $G$ is the gravitational constant, $m_{\rm p}$ is the mass of the proton, $c$ is the speed of light, and $\sigma_{\rm T}$ is the Thomson cross-section. The bolometric luminosity ($L_{\rm Bol}$) of the AGN in our sample was calculated from the intrinsic 2--10\,keV luminosity, using a 2--10\,keV bolometric correction of $\kappa_{2-10}=20$ (\citealp{Vasudevan:2009ng}; $L_{\rm Bol}=\kappa_{2-10}\times L_{2-10}$). In $\S$\ref{sec:Ecvsaccretion} we discuss the effects of considering a dependence of $\kappa_{2-10}$ on $L_{\rm Bol}$ and/or $\lambda_{\rm Edd}$ (e.g., \citealp{Vasudevan:2009ng}). The typical uncertainty on $\lambda_{\rm Edd}$ is conservatively estimated to be $\sim 0.5$\,dex (see \citealp{Koss:2017fp}).


\section{The relation between the cutoff energy and the physical properties of the accreting SMBH}\label{sec:Ecvsaccretion}

Using the BASS database we explored the relation between the cutoff energy and the properties of the accreting SMBH, such as its luminosity, black hole mass and Eddington ratio. 
In the left panel of Fig.\,\ref{fig:Ecut_vsL} we show the cutoff energy versus the 14-150\,keV intrinsic (absorption and K-corrected) luminosity ($L_{14-150}$). Since the sample contains a large number of lower limits, we used the Kaplan-Meier estimator within the \textsc{asurv} package \citep{Feigelson:1985qv,Isobe:1986ys}, using a \textsc{python} implementation (see \S5 of \citealp{Shimizu:2016hc} for details) to calculate the median values of $E_{\rm C}$ in several luminosity bins. As shown in the right panel of Fig.\,\ref{fig:Ecut_vsL}, the sample does not show significant changes of the cutoff energy with the AGN luminosity.
We performed a linear fit on the binned data, which include the censored values, using a relation of the form $E_{\rm C}=\alpha+\beta\log L_{14-150}$. The p-value of the correlation is $0.74$, suggesting that no significant trend exists between the cutoff energy and the intrinsic 14--150\,keV AGN luminosity.

In the left panel of Fig.\,\ref{fig:Ecut_vsMBH} we plot the cutoff energy versus $M_{\rm BH}$ for the 211 unobscured AGN for which this parameter is available. The rebinned plot (right panel of Fig.\,\ref{fig:Ecut_vsMBH}), shows  a positive trend. The median $E_{\rm C}$ appears to increase with $M_{\rm BH}$, from $123\pm50$\,keV for $5 \leq \log (M_{\rm BH}/M_{\odot})<6$ to $323\pm51$\,keV for $9 \leq \log (M_{\rm BH}/M_{\odot})<10$. Fitting the data with $E_{\rm C}=\gamma+\delta\log M_{\rm BH}$, we found  a significant correlation, with a p-value of $0.003$ and a slope of $\delta=49\pm16$. A similar trend is observed when considering only the objects for which the black hole mass was estimated using broad H$\beta$ (blue line in Fig.\,\ref{fig:Ecut_vsMBH}).

The left panel of Fig.\,\ref{fig:Ecut_vsEddratio} shows the scatter plot of $E_{\rm C}$ versus Eddington ratio. The rebinned plot (right panel of Fig.\,\ref{fig:Ecut_vsEddratio}) shows a negative trend, with a clear difference in the typical $E_{\rm C}$ for objects accreting at high and low Eddington ratio: the median cutoff energy of the AGN accreting at $\lambda_{\rm Edd}\leq 0.1$ is $E_{\rm C}=370\pm51$\,keV, while the sources at $0.1 < \lambda_{\rm Edd}\leq 1$ have a median of $E_{\rm C}=160\pm41$\,keV, which implies a $3.2\sigma$ difference between the two subsets of sources. Such a difference is confirmed also considering only the closest AGN ($z\leq 0.05$): for $\lambda_{\rm Edd}\leq 0.1$ we find $E_{\rm C}=506\pm82$\,keV, while for $0.1 < \lambda_{\rm Edd}\leq 1$ the median cutoff energy is $E_{\rm C}=164\pm46$\,keV (i.e., the difference between the subsamples is $3.6\sigma$). Ignoring the lower limit on $E_{\rm C}$ results in a rather flat trend, and no significant difference in $E_{\rm C}$ is found between objects accreting at low and high $\lambda_{\rm Edd}$.
Fitting the rebinned data with $E_{\rm C}=\epsilon+\zeta\log \lambda_{\rm Edd}$, we obtained a p-value of $0.01$, and a slope of $\zeta=-74\pm31$.

\begin{figure}
\centering
\includegraphics[width=0.48\textwidth]{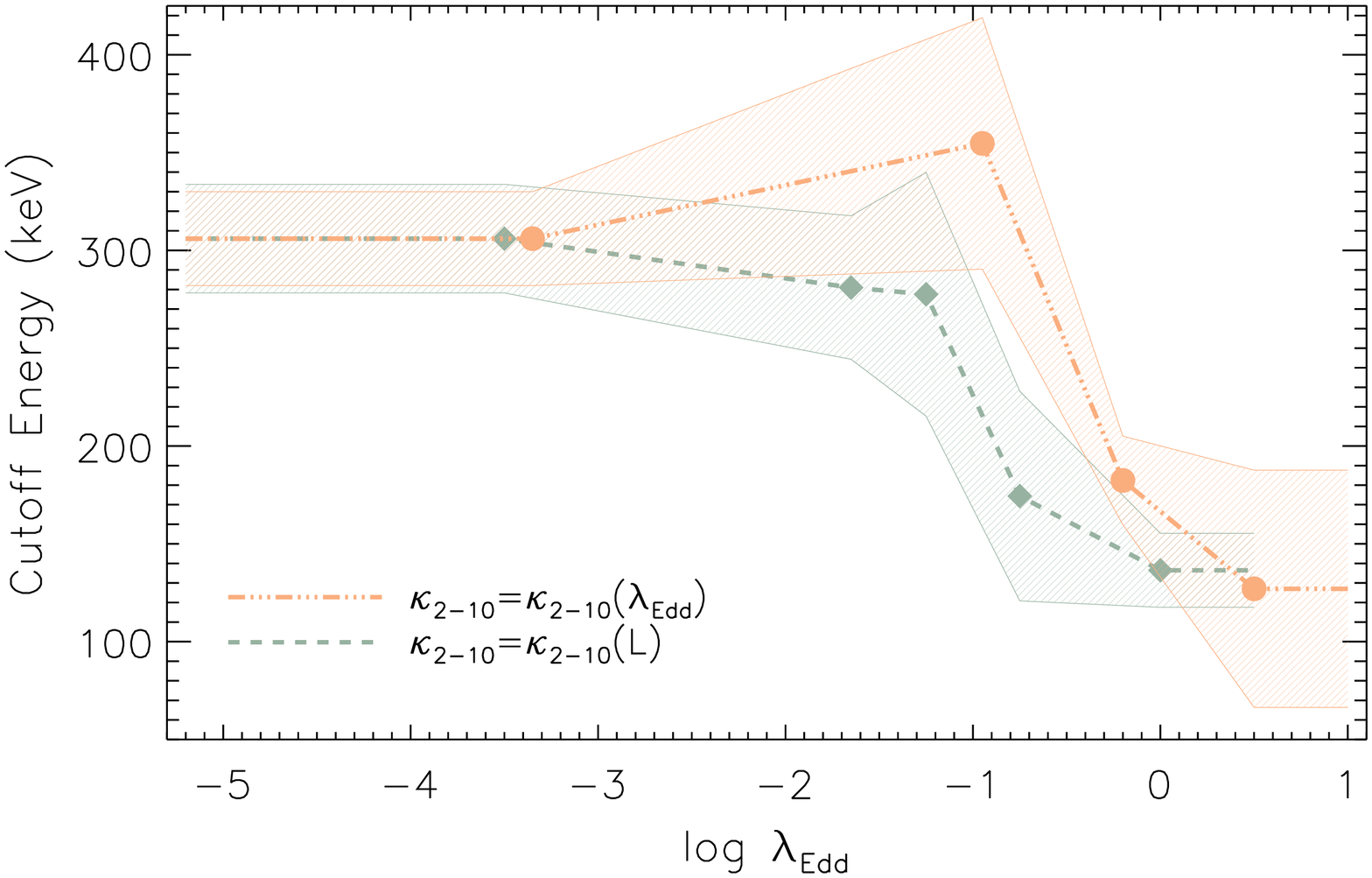}
  \caption{Median values of the cutoff energy for different intervals of Eddington ratio considering different 2--10\,keV bolometric corrections: the $\lambda_{\rm Edd}$-dependent bolometric corrections of \citet{Vasudevan:2007qt} (orange dot-dot-dashed line) and the luminosity-dependent bolometric corrections of \citet{Lusso:2012it} (green dashed line). The shaded area corresponds to the median absolute deviation.}
\label{fig:Ecut_edd_BC}
\end{figure}

In Fig.\,\ref{fig:Ecut_edd_BC} we illustrate the effect of adopting $\lambda_{\rm Edd}$-dependent (\citealp{Vasudevan:2007qt}, orange dot-dot-dashed line) and luminosity-dependent (\citealp{Lusso:2012it}, green dashed line) 2--10\,keV bolometric corrections to the relation between $E_{\rm C}$ and $\lambda_{\rm Edd}$. In both cases we find the same trend observed adopting a constant $\kappa_{2-10}=20$: objects accreting at higher Eddington ratios tend to have lower cutoff energies. In particular we find that, using the corrections of \citet{Vasudevan:2007qt}, the median cutoff energy drops from $E_{\rm C}=342\pm27$\,keV for $\log\lambda_{\rm Edd} \leq -0.7$ to $E_{\rm C}=163\pm45$\,keV for $\log\lambda_{\rm Edd} > -0.7$, implying a difference significant at the $3.4\sigma$ level. Considering the corrections of \citet{Lusso:2012it} the difference is of $2.8\sigma$ ($E_{\rm C}=359\pm54$\,keV for $\log\lambda_{\rm Edd} \leq -1$ and $E_{\rm C}=160\pm45$\,keV for $\log\lambda_{\rm Edd} > -1$).

\begin{figure*}
\centering
\includegraphics[width=0.48\textwidth]{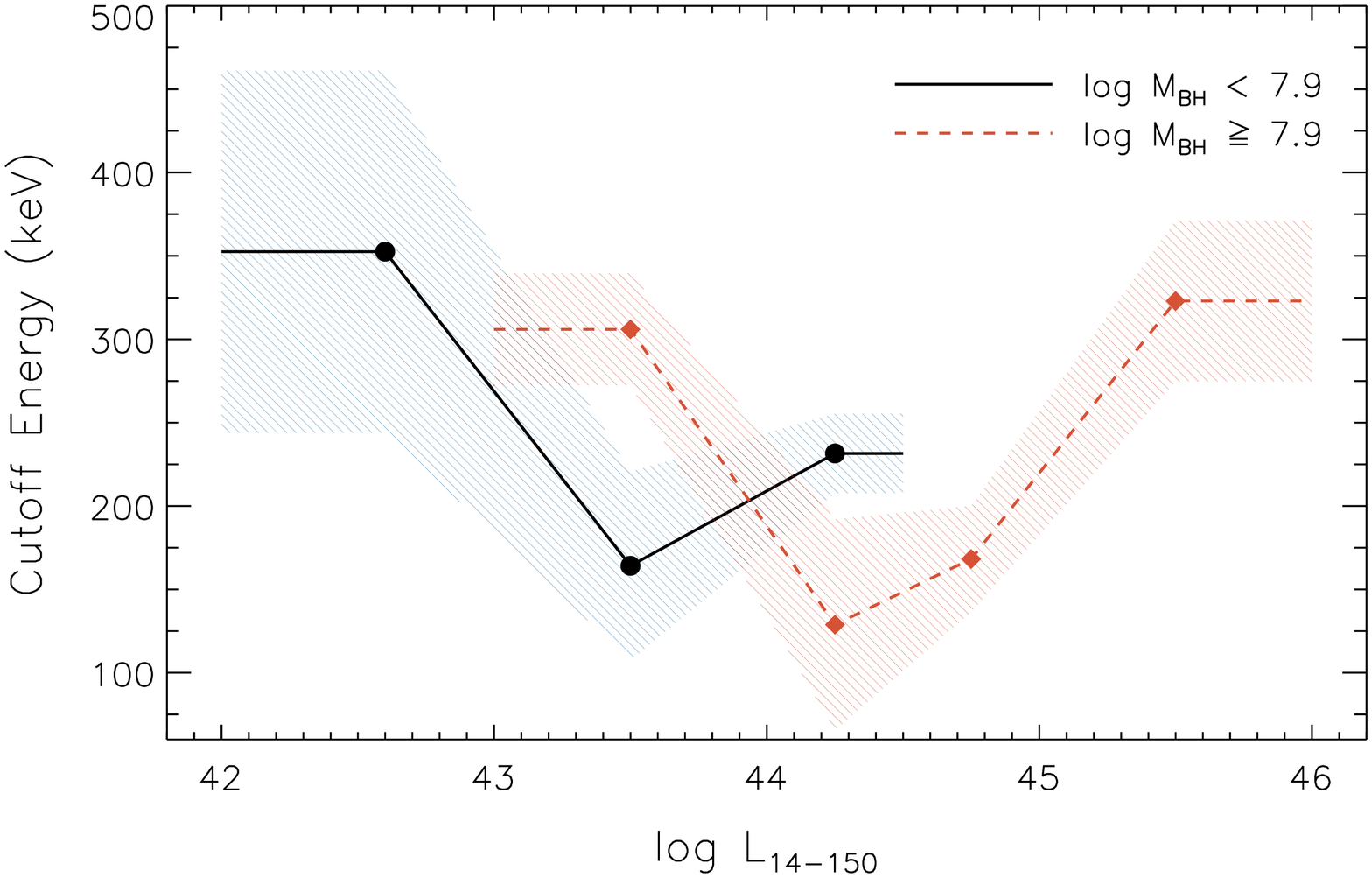}
\includegraphics[width=0.48\textwidth]{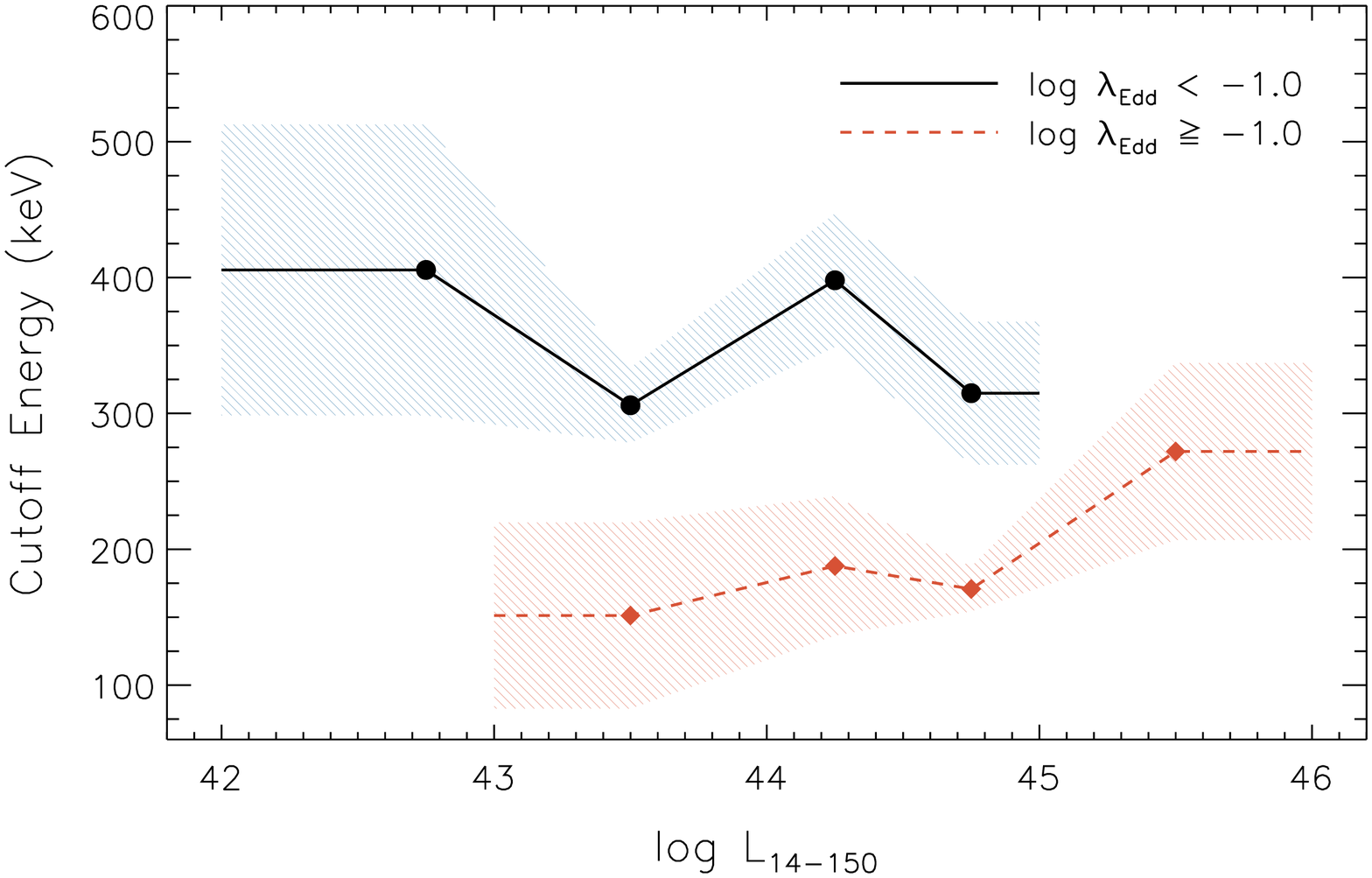}
  \caption{Cutoff energy versus the luminosity for two ranges of black hole mass ({\it left panel}, in units of $M_{\odot}$) and of Eddington ratio ({\it right panel}). Both panels show the median values of the cutoff energy for different intervals of $L_{14-150}$ (in $\rm erg\,s^{-1}$). The shaded areas corresponds to the median absolute deviations. The figures show little or no dependence of $E_{\rm C}$ on the luminosity, while there is a clear difference between sources accreting at different Eddington ratios: the AGN with $\lambda_{\rm Edd}\geq 0.1$ tend to have lower cutoff energies than those with $\lambda_{\rm Edd}< 0.1$, even at similar luminosities.}
\label{fig:Ecut_vsLum_othPar}
\end{figure*}

\begin{figure*}
\centering
\includegraphics[width=0.48\textwidth]{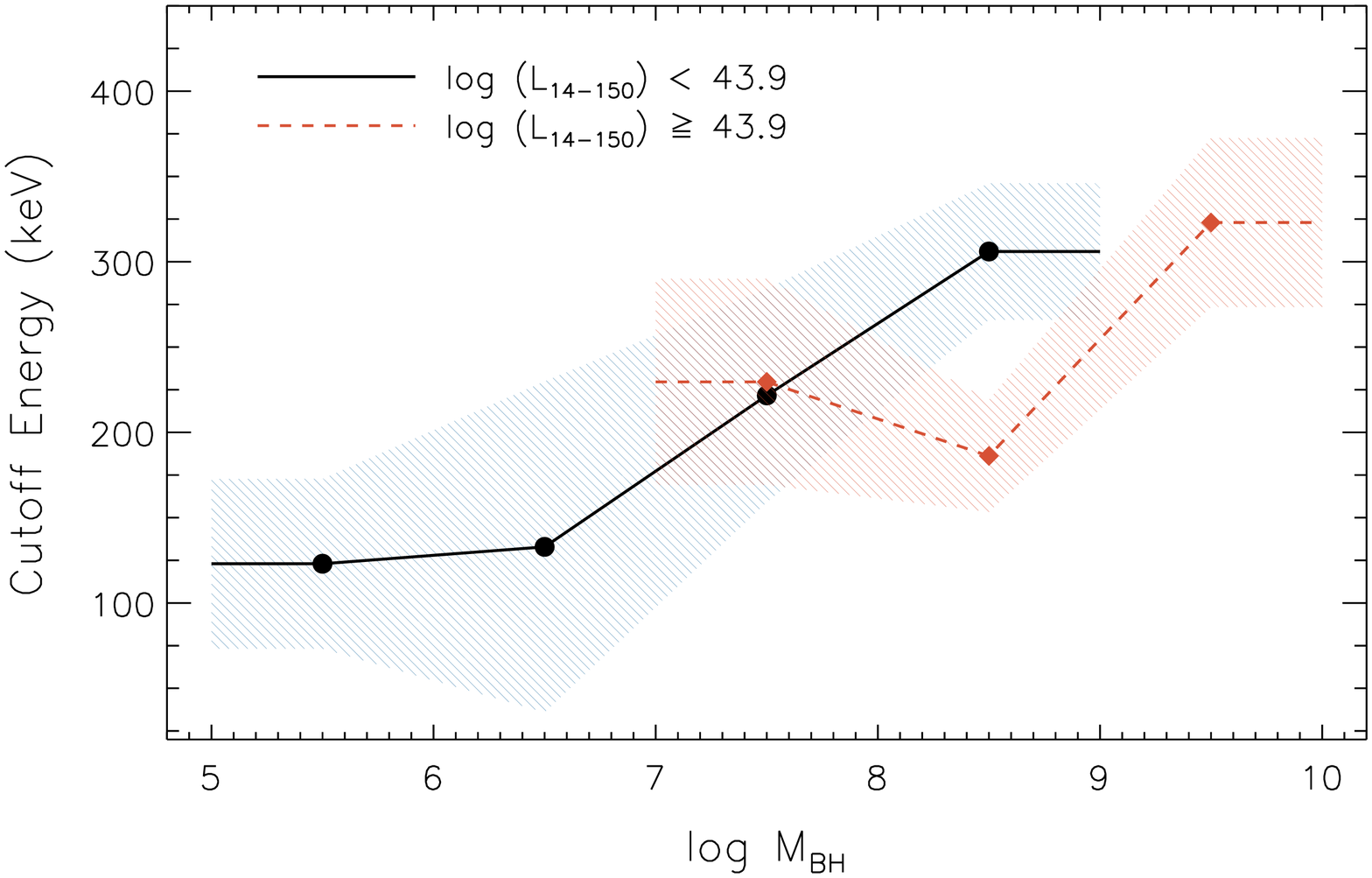}
\includegraphics[width=0.48\textwidth]{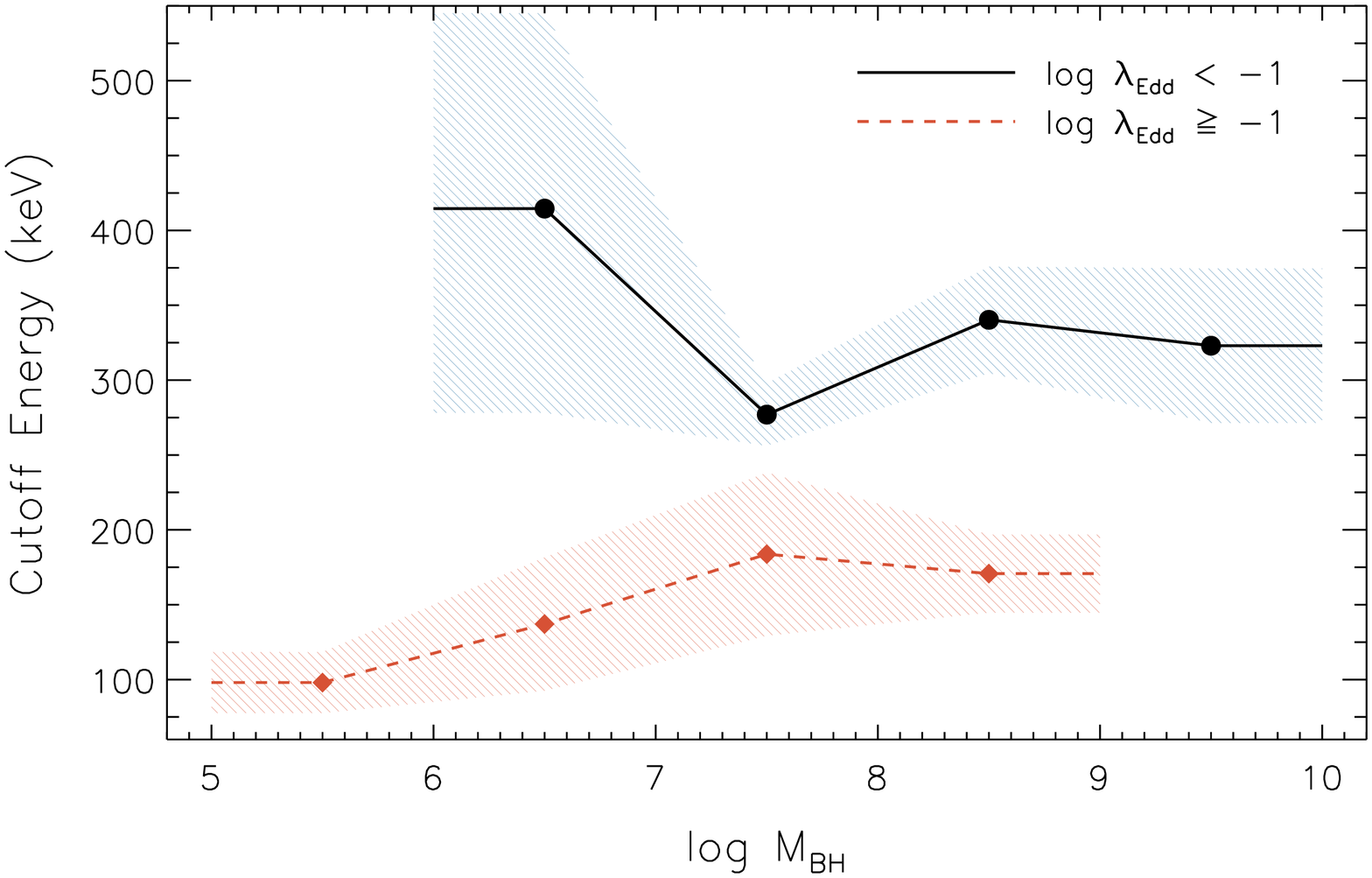}
  \caption{Cutoff energy versus the $M_{\rm\,BH}$ (in $M_{\odot}$) for two intervals of luminosity ({\it left panel}, in $\rm erg\,s^{-1}$) and Eddington ratio ({\it right panel}). Both panels show the median cutoff energies; the shaded areas corresponds to the median absolute deviations. The figures illustrate how the dependence of $E_{\rm C}$ on black hole mass disappears when dividing the sample into bins of Eddington ratio, and that sources with $\lambda_{\rm Edd}\geq 0.1$ tend to have lower cutoff energies than those with $\lambda_{\rm Edd}< 0.1$, regardless of the interval of $M_{\rm BH}$.}\label{fig:Ecut_vsMBH_othPar}. 
\end{figure*}

\begin{figure*}
\centering
\includegraphics[width=0.48\textwidth]{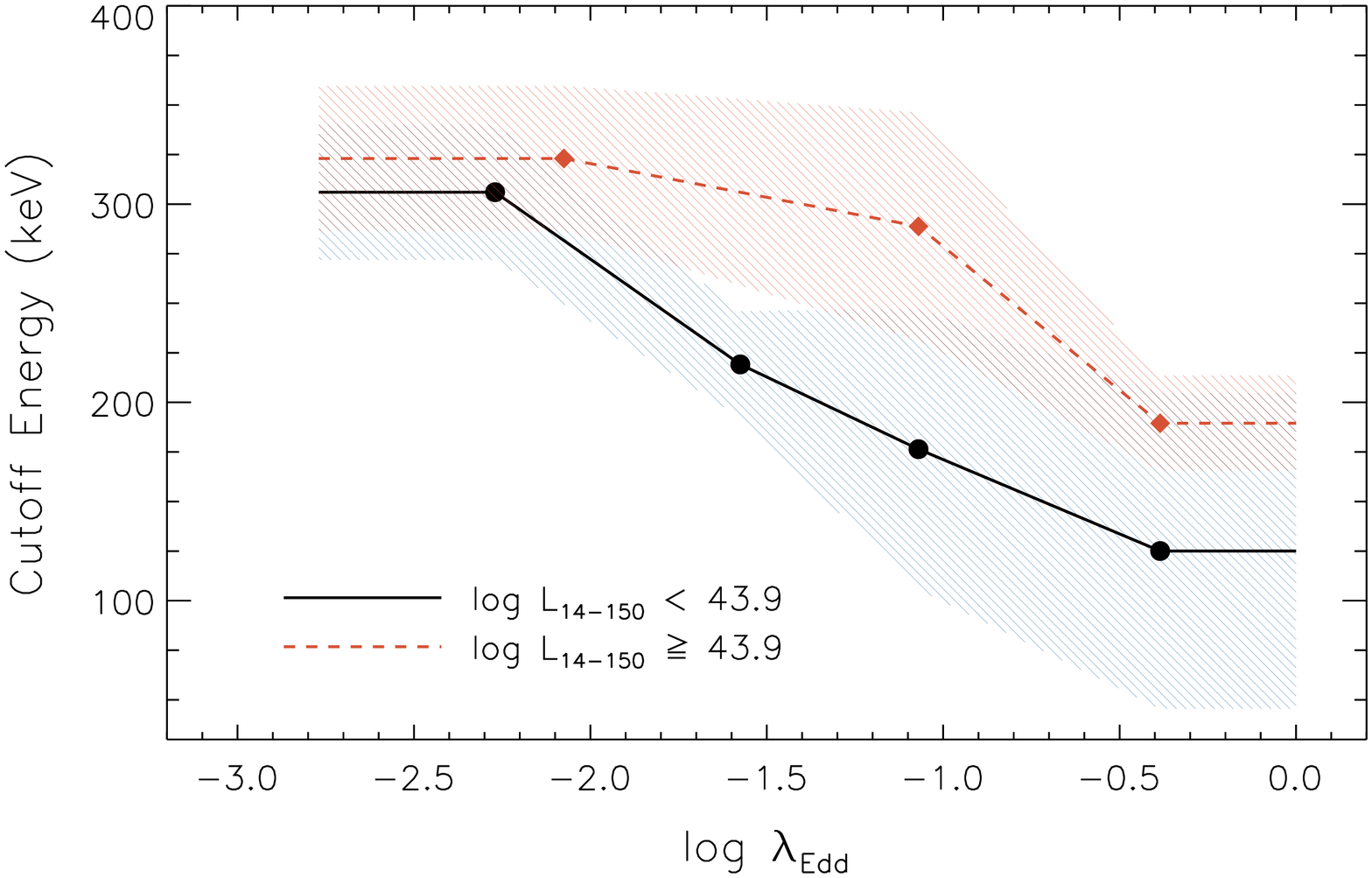}
\includegraphics[width=0.48\textwidth]{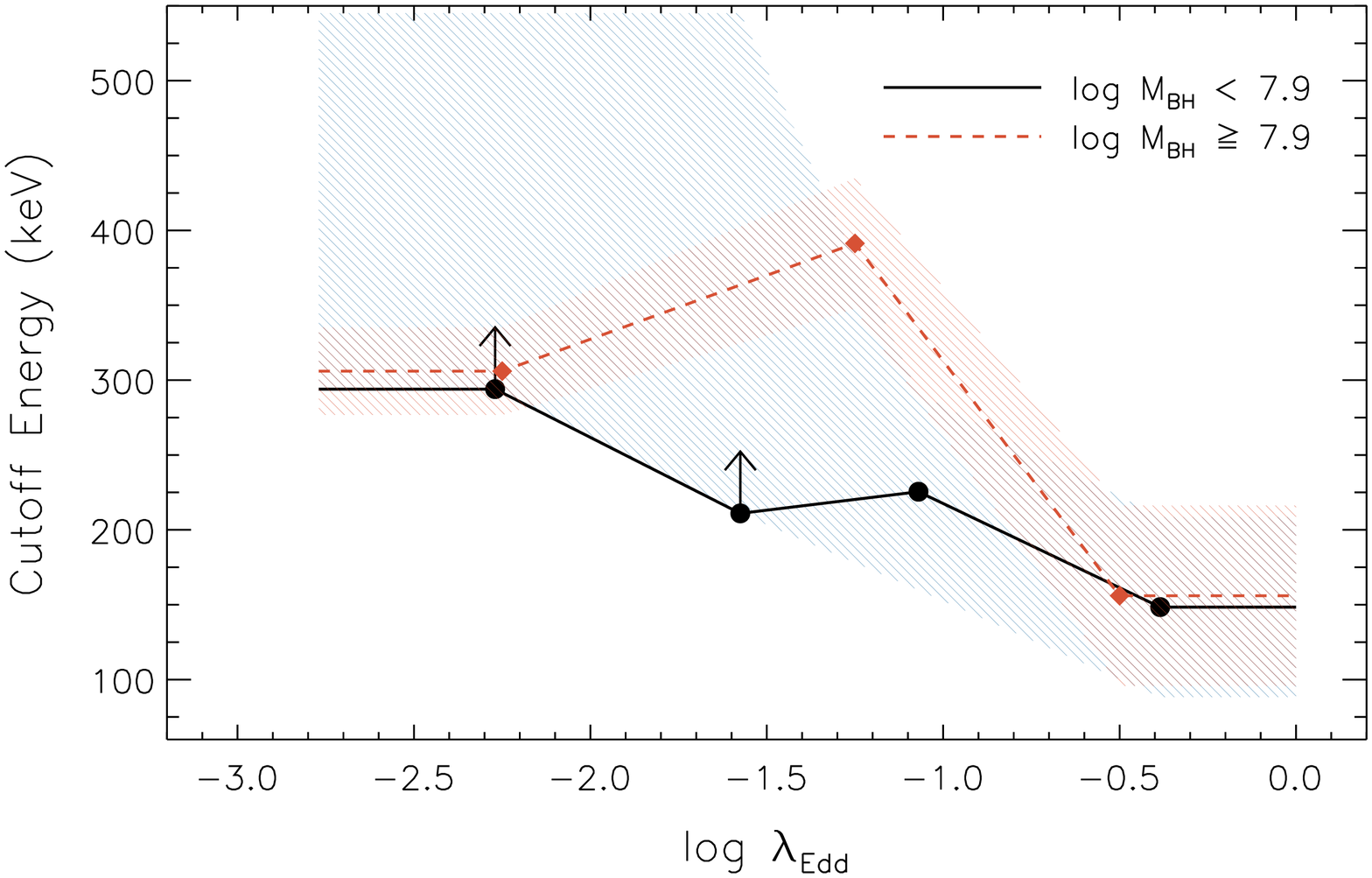}
  \caption{Cutoff energy versus the Eddington ratio for two ranges of luminosity ({\it left panel}, in $\rm erg\,s^{-1}$) and black hole mass ({\it right panel}, in $M_{\odot}$). Both panels show the median cutoff energies; the shaded areas corresponds to the median absolute deviations. In the right panel the first two bins of $\log (M_{\rm BH}/M_{\odot})< 7.9$ are lower limits because only censored data are available in that interval of $\lambda_{\rm Edd}$ and $M_{\rm BH}$. The plots show that sources with $\lambda_{\rm Edd}\geq 0.1$ tend to have lower cutoff energies than those with $\lambda_{\rm Edd}< 0.1$, regardless of the black hole mass or luminosity, thus confirming that the Eddington ratio is the main physical parameter controlling $E_{\rm C}$. }\label{fig:Ecut_vsEddratio_othPar}
\end{figure*}

\begin{figure*}
\centering
\includegraphics[width=0.48\textwidth]{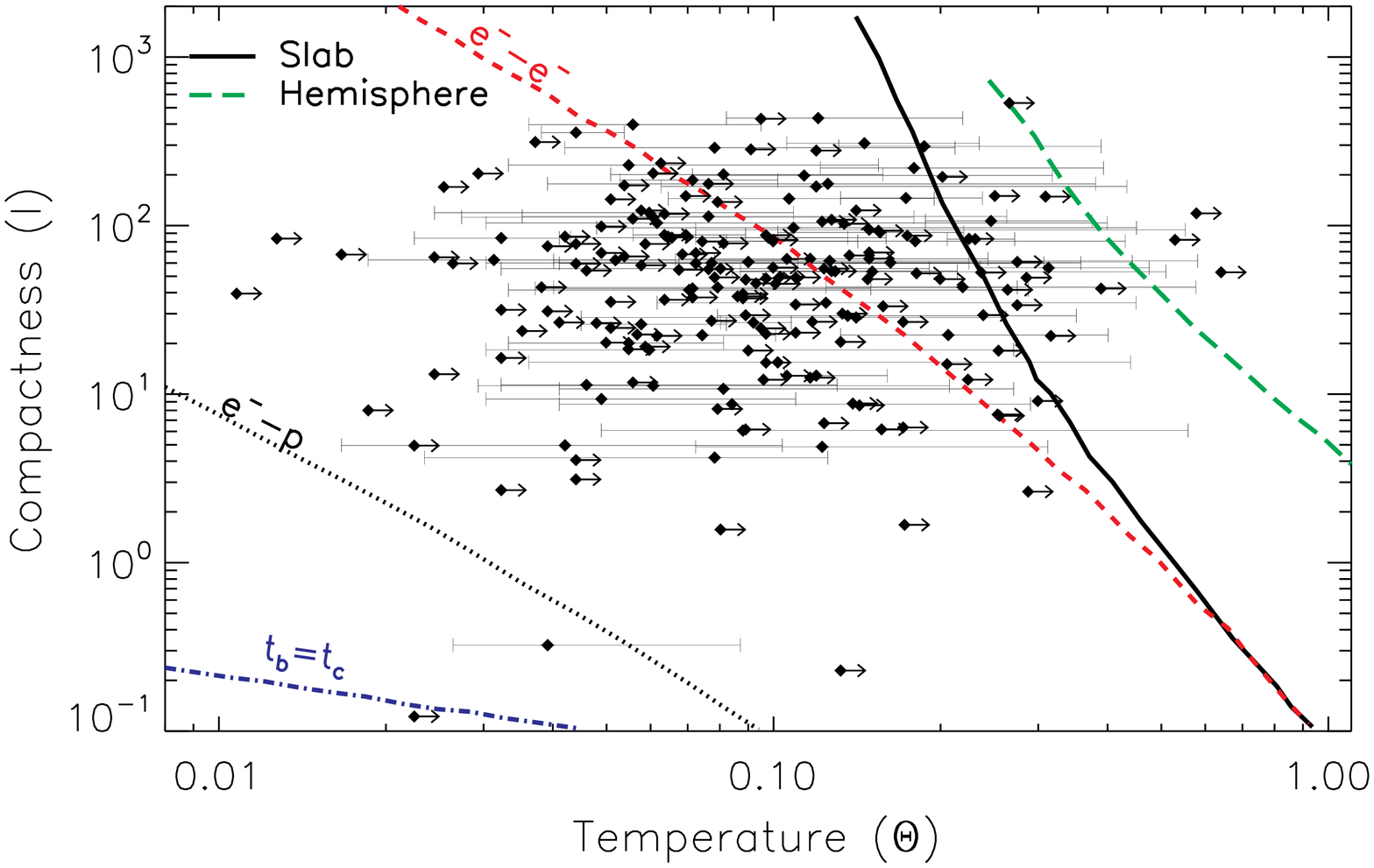}
\includegraphics[width=0.48\textwidth]{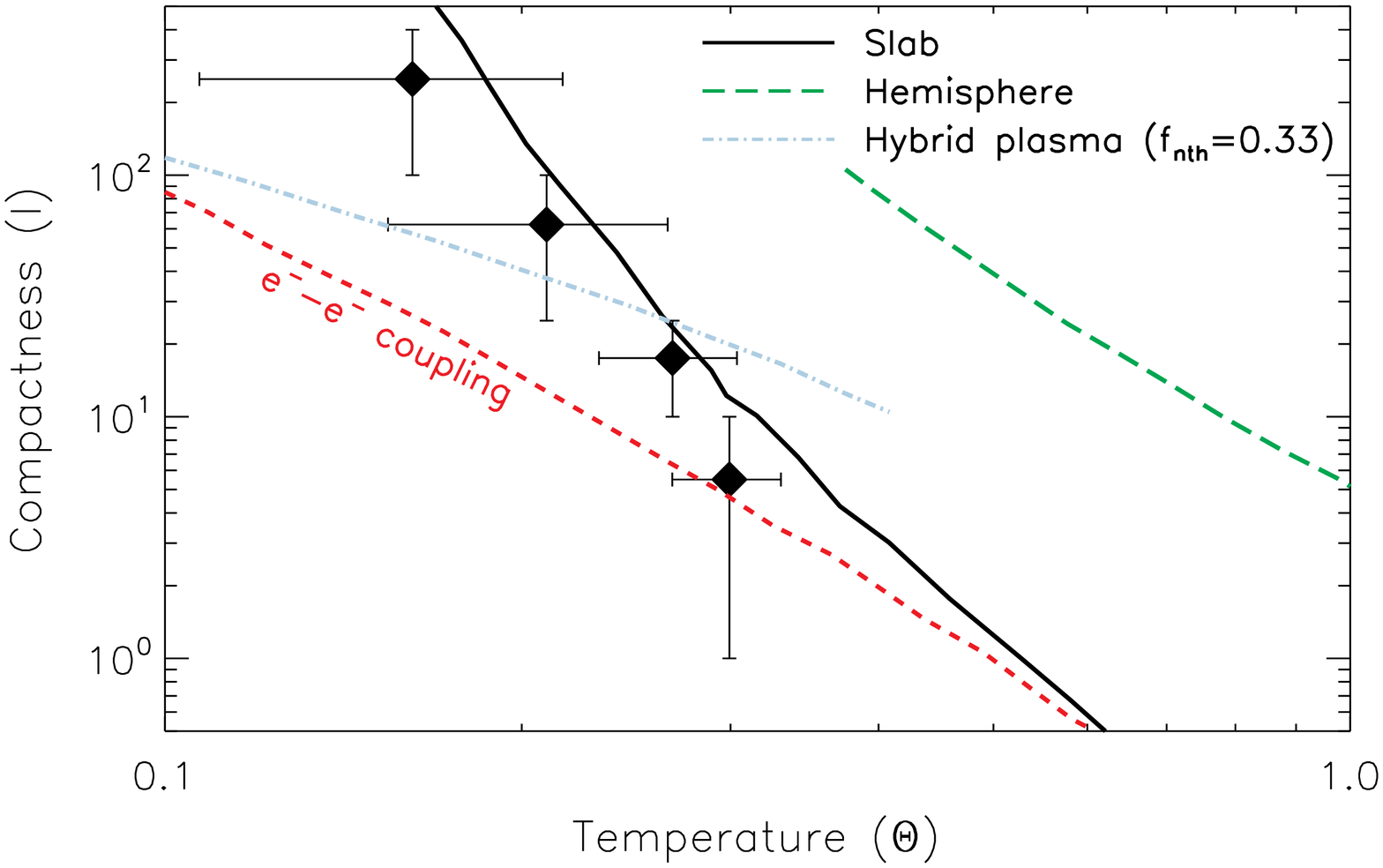}
  \caption{{\it Left panel:} Compactness-temperature diagram for the 211 AGN in our hard X-ray selected sample for which cutoff energies \citep{Ricci:2017bf} and black hole masses \citep{Koss:2017fp} were available within BASS. The blue dot-dashed curve shows the limit of the region where bremsstrahlung dominates, while the black dotted and red dashed curves show the boundary to the region dominated by electron-proton and electron-electron coupling, respectively.
  The continuous black and the long dashed green curves represent the runaway pair production limits for a slab and a hemisphere corona \citep{Stern:1995sh}. 
  {\it Right panel:} same as left panel, but showing the median of the temperature parameter, obtained including the lower limits using the Kaplan-Meier estimator.  The cyan dot-dot-dashed curve shows the runaway pair production limit obtained considering a hybrid plasma with 33\% of non-thermal electrons \citep{Fabian:2017dq}. }
\label{fig:Ecut_thetal}
\end{figure*}

To further test the relation between $E_{\rm C}$, $M_{\rm BH}$ and $\lambda_{\rm Edd}$, we used a different approach to calculate the median values of the cutoff energy for different values of black hole mass and Eddington ratio. This was done performing 1,000 Monte Carlo simulations for each object of our sample, substituting the cutoff energies we could constrain with values that were randomly selected from a Gaussian distribution centered on $E_{\rm C}$, and with a standard deviation given by the uncertainty. Lower limits ($LL$) were substituted with values randomly selected from a uniform distribution in the interval [$LL$, $E_{\rm C}^{\rm\,max}$], where $E_{\rm C}^{\rm\,max}=1000$\,keV is the maximum cutoff energy. For each Monte Carlo run we calculated the median in two different bins of $M_{\rm BH}$ and $\lambda_{\rm Edd}$, and finally we calculated the means of all simulations. For $10^{5}\leq M_{\rm BH}/M_{\odot}< 10^{7.5}$ we obtain $E_{\rm C}=312\pm44$\,keV, while for $10^{7.5}\leq M_{\rm BH}/M_{\odot}< 10^{10}$ we found $E_{\rm C}=416\pm30$\,keV. This implies a $\simeq 2\sigma$ difference between the two subsamples.
For $\lambda_{\rm Edd}\leq 0.1$ we find $E_{\rm C}=432\pm30$\,keV, while for $0.1 < \lambda_{\rm Edd}\leq 1$ the median cutoff energy is $E_{\rm C}=307\pm37$\,keV. This implies a difference significant at the $2.6\sigma$ level. It should be remarked that the median values obtained using this approach are typically larger than those we found using the survival analysis. This is due to the fact that we are assuming a flat distribution for the lower limits, which likely does not represent the real physical distribution of plasma temperatures, and largely increases the number of objects with $E_{\rm C}>500$\,keV.

To investigate whether the Eddington ratio or the black hole mass is the main physical parameter responsible for differences in the cutoff energy, in Figs.\,\ref{fig:Ecut_vsLum_othPar}--\ref{fig:Ecut_vsEddratio_othPar} we plot the median values of $E_{\rm C}$ as a function of luminosity, black hole mass and Eddington ratio. In each of the six panels we illustrate the dependence on one of these parameters for two subsets of sources covering different intervals of the other parameters. No clear dependence of $E_{\rm C}$ on the X-ray luminosity is found dividing the sample into bins of $M_{\rm BH}$ and $\lambda_{\rm Edd}$ (left and right panels of Fig.\,\ref{fig:Ecut_vsLum_othPar}, respectively), although a difference can be observed between the low and high Eddington ratio subsamples. Interestingly, while a possible trend between $E_{\rm C}$ and $M_{\rm BH}$ is observed dividing the sample in two luminosity intervals (left panel of Fig.\,\ref{fig:Ecut_vsMBH_othPar}), such a relation disappears when splitting the sources into bins of Eddington ratio (right panel of Fig.\,\ref{fig:Ecut_vsMBH_othPar}). A similar trend is observed considering only objects for which $M_{\rm BH}$ was obtained from H$\beta$. This, together with the fact that the subsample with $\lambda_{\rm Edd}\leq 0.1$ has a lower median $E_{\rm C}$ than that with $0.1 < \lambda_{\rm Edd}\leq 1$ across the interval of black hole masses spanned by the data suggests that the correlation between $E_{\rm C}$ and $\lambda_{\rm Edd}$ is the main relation. This is confirmed by the fact that, regardless of the luminosity (left panel of Fig.\,\ref{fig:Ecut_vsEddratio_othPar}) and black hole mass (right panel of Fig.\,\ref{fig:Ecut_vsEddratio_othPar}), sources with high $\lambda_{\rm Edd}$ tend to have lower cutoff energies than those with low mass-normalised accretion rates.


\section{AGN in the temperature--compactness parameter space}\label{sec:thetalplane}

Two important parameters of AGN coronae are their compactness \citep{Cavaliere:1980cs,Guilbert:1983ek} and normalised temperature. The compactness parameter ($l$) is defined as
\begin{equation}\label{eq:l}
l=\frac{L_{\rm X}}{R_{\rm X}}\frac{\sigma_{\rm T}}{m_{\rm e}c^3}=4\pi\frac{\lambda_{\rm Edd}}{\kappa_{\rm x}}\frac{m_{\rm p}}{m_{e}}\frac{R_{\rm g}}{R_{\rm X}},
\end{equation}
where $L_{\rm X}$ is the X-ray luminosity of the source, $R_{\rm X}$ is the radius of the X-ray source, $\kappa_{\rm x}$ is the X-ray bolometric correction, $m_{\rm e}$ is the mass of the electron, $m_{\rm p}$ is the mass of the proton, and $\sigma_{\rm T}$ is the Thomson cross-section. 
The compactness was calculated using the 0.1--200\,keV luminosity, which was obtained from the intrinsic 14--150\,keV luminosity, assuming $\Gamma=1.8$ \citep{Mushotzky:1982rp,Winter:2009xi,Ricci:2017bf}, while the 0.1--200\,keV bolometric correction was set to $\kappa_{\rm x}=3.87$, which corresponds to our assumption of $\kappa_{2-10}=20$ \citep{Vasudevan:2009ng} and the same $\Gamma$. The normalised temperature parameter ($\Theta$) is:
\begin{equation}\label{eq:theta}
\Theta=\frac{kT_{\rm e}}{m_{\rm e}c^2}=\frac{E_{\rm C}}{2m_{\rm e}c^2}.
\end{equation}
In the above equation we considered that $kT_{\rm e}=E_{\rm C}/2$, which is an approximation valid for optically-thin plasma\footnote{As discussed in \citet{Petrucci:2001nq}, for $\tau \gg 1$ then $kT_{\rm e}\simeq E_{\rm C}/3$.} for a corona with slab geometry \citep{Petrucci:2000kq,Petrucci:2001nq}, obtained using the Comptonization model of \citet{Haardt:1994ys}. Considering this relation, the median temperature of the X-ray emitting plasma for the objects of our sample is $kT_{\rm e}=105\pm18$\,keV.

In the left panel of Fig.\,\ref{fig:Ecut_thetal} we show the temperature--compactness diagram for the {\it Swift}/BAT AGN in our sample for which black hole masses are available, assuming $R_{\rm X}=10R_{\rm g}$. Several regions can be defined in this diagram, depending on the process dominating the electron cooling (see \citealp{Fabian:2015db} and references therein for a detailed discussion). The region where bremsstrahlung dominates the cooling of electrons is defined by $l\lesssim 3\alpha_{\rm f}\Theta^{-1/2}$ (blue dot-dashed curve), where $\alpha_{\rm f}$ is the fine-structure constant. Electron-proton and electron-electron collisions occur faster than the electron cooling for compactness and temperatures lower than the values delimited by the black dotted curve and the red dashed curve, respectively \citep{Ghisellini:1993fj,Fabian:1994ht}.

Pair production, due to photon-photon collisions, can be a fundamental process in coronae \citep{Svensson:1982rw,Svensson:1982eu,Guilbert:1983ek}. This process could lead to runaway pair production, acting as a thermostat for the corona \citep{Bisnovatyi-Kogan:1971nr,Svensson:1984gs,Zdziarski:1985tg,Fabian:2015db,Fabian:2017dq}. 
The region where there is runaway pair production is delimited by the {\it pair line}, which is illustrated as a green dashed curve (following \citealp{Svensson:1984gs}) and as a black continuous curve (following \citealp{Stern:1995sh}) in Fig.\,\ref{fig:Ecut_thetal} for an isolated cloud and for a slab corona, respectively. If an X-ray source moves into this region of the parameter space (by an increase in its temperature or compactness), then it starts to rapidly form pairs, which increases the number of particles sharing the available power, causing the energy per particle (i.e., the temperature) to drop. {\it Sources are therefore expected to typically lie at the edge of the pair region.}

\begin{figure}
\centering
\includegraphics[width=0.48\textwidth]{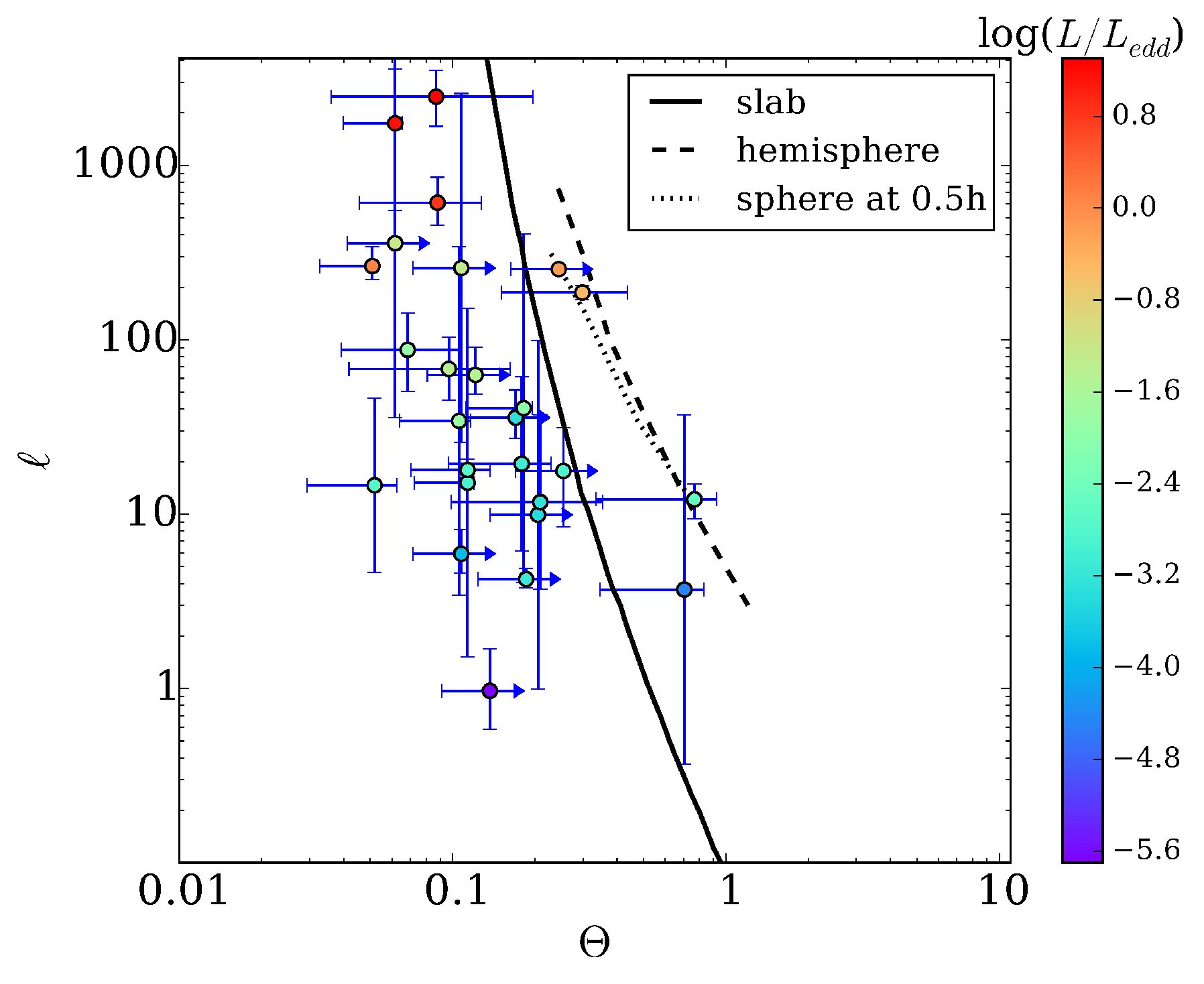}
  \caption{Temperature ($\Theta$)-compactness ($l$) diagram for the AGN from \citet{Fabian:2015db}, color-coded according to their Eddington ratio. The cutoff energy were inferred using {\it NuSTAR} observations. The continuous, dashed and dotted lines represent the pair lines for different geometries of the corona: a slab, a hemisphere, and a sphere at a height equal to half the radius of the sphere, respectively. Consistently with what we found for our sample, objects at low temperature and high compactness tend to have higher $\lambda_{\rm Edd}$ than those at high temperature and low compactness.}
\label{fig:theta_l_nustar}
\end{figure}

The right panel of Fig.\,\ref{fig:Ecut_thetal} shows the median values of $\Theta$ in different bins of $l$. The medians were calculated using the Kaplan-Meier estimator, including the lower limits, as discussed in \S\ref{sec:Ecvsaccretion}. We also show the pair line for a hybrid plasma with 33\% of the electrons being non-thermal \citep{Fabian:2017dq}. The plot illustrates that, in general, AGN are concentrated close to the pair line corresponding to a slab, avoiding the runaway pair production region, in agreement with theoretical predictions. Since plasmas are expected to concentrate right on the edge of the relevant pair-production regions in the compactness-temperature parameter space (see above), this suggests that the shape of the X-ray corona might be better approximated as a slab rather than sphere. This would also easily explain the observed dependence of the cutoff energy on the Eddington ratio. For a fixed value of $R$ (in $R_{\rm g}$), $\l$ is in fact directly proportional to the Eddington ratio ($l\propto\frac{\lambda_{\rm Edd}R_{\rm g}}{\kappa_{\rm x}R}$, see Eq.\,\ref{eq:l}), and since $\Theta$ decreases with $\l$, one would expect that AGN accreting at high $\lambda_{\rm Edd}$ would also tend to have X-ray emitting plasma with lower temperatures. This is also consistent with what is found for the AGN from \citet{Fabian:2015db} (Fig.\,\ref{fig:theta_l_nustar}): objects at low temperature and high compactness tend to have higher $\lambda_{\rm Edd}$ (e.g., Ark\,564, see \citealp{Kara:2017lq}) than those at high temperature and low compactness.

\begin{figure}
\centering
\includegraphics[width=0.48\textwidth]{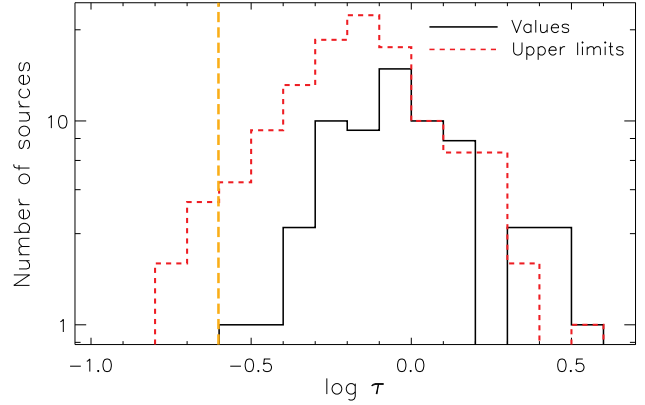}
  \caption{Histogram of the optical depth of the Comptonizing plasma, calculated using Eq.\,\ref{eq:tau_gamma_kte} (see \S\ref{sect:tau}). The continuous black and dashed red lines illustrate the values and the upper limits, respectively. The vertical dashed orange line shows the median of the sample ($\tau=0.25 \pm 0.06$), calculated taking into account the upper limits.}
\label{fig:tau_histo}
\end{figure}

\section{The plasma optical depth and its relation with the accretion properties of AGN}\label{sect:tau}

In this section we explore the relation between the optical depth of the Comptonizing plasma and the properties of the accreting SMBH. While $\tau$ is not a parameter directly obtained by our broad-band X-ray spectral analysis, it can be constrained indirectly using the dependence of $\Gamma$ on $kT_{\rm e}$ and $\tau$. The photon index decreases for increasing values of the Compton parameter ($y$; e.g., \citealp{Rybicki:1979wd}), which is defined as:
\begin{equation}
y = \max(\tau, \tau^2)\times \frac{4kT_{\rm e}}{m_{\rm e} c^2}.
\end{equation}

\begin{figure*}
\centering
\includegraphics[width=0.48\textwidth]{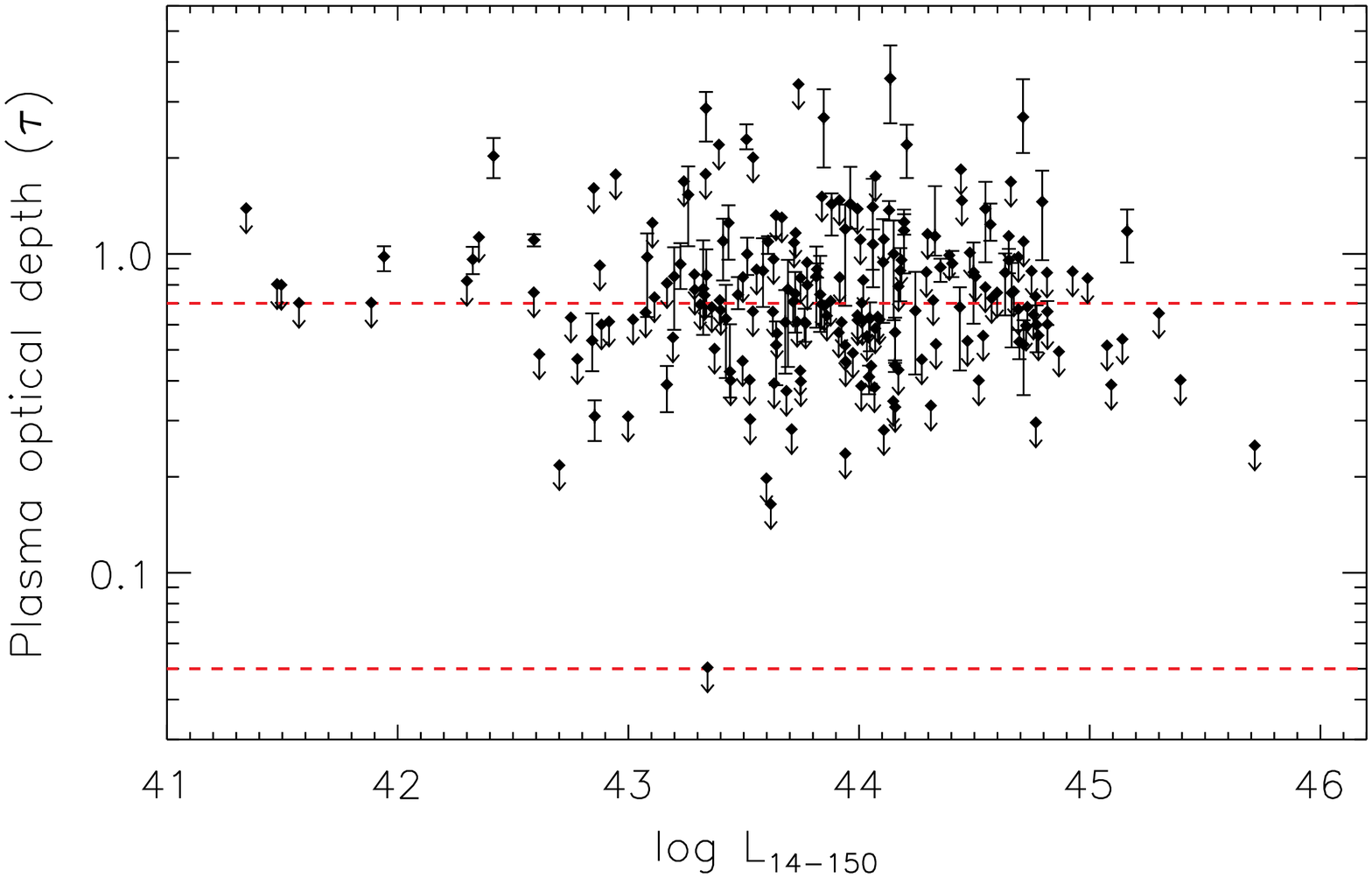}
\includegraphics[width=0.48\textwidth]{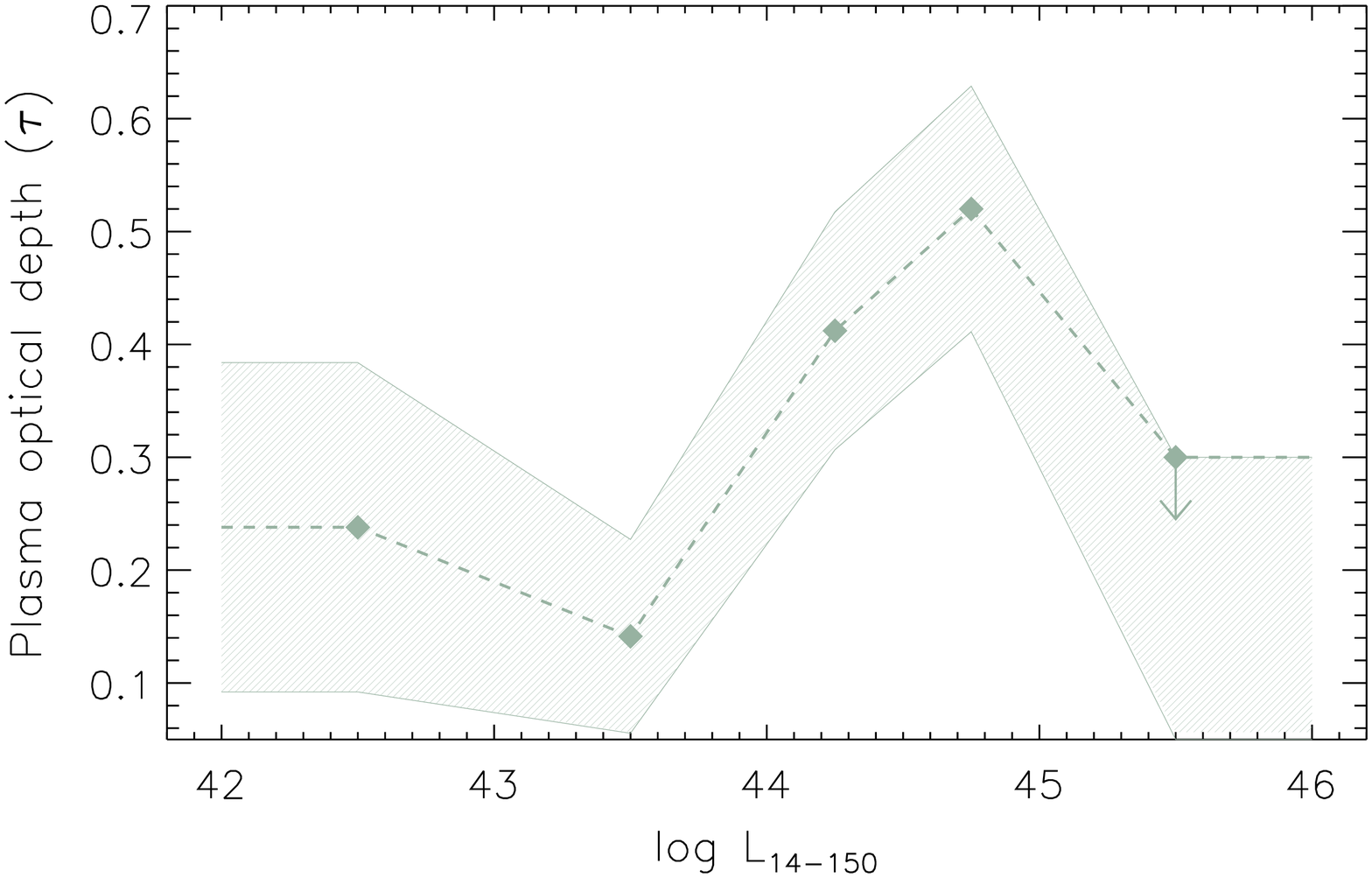}
  \caption{{\it Left panel:} Optical depth versus the 14--150\,keV intrinsic luminosity (in $\rm erg\,s^{-1}$). The red dashed lines show the interval of $\tau$ shown in the right panel. {\it Right panel:} Median of $\tau$ for different intervals of $L_{\rm 14-150}$, calculated including the lower limits using the Kaplan-Meier estimator. The shaded area corresponds to the median absolute deviation.}
\label{fig:tau_vsL}
\end{figure*}

\begin{figure*}
\centering
\includegraphics[width=0.48\textwidth]{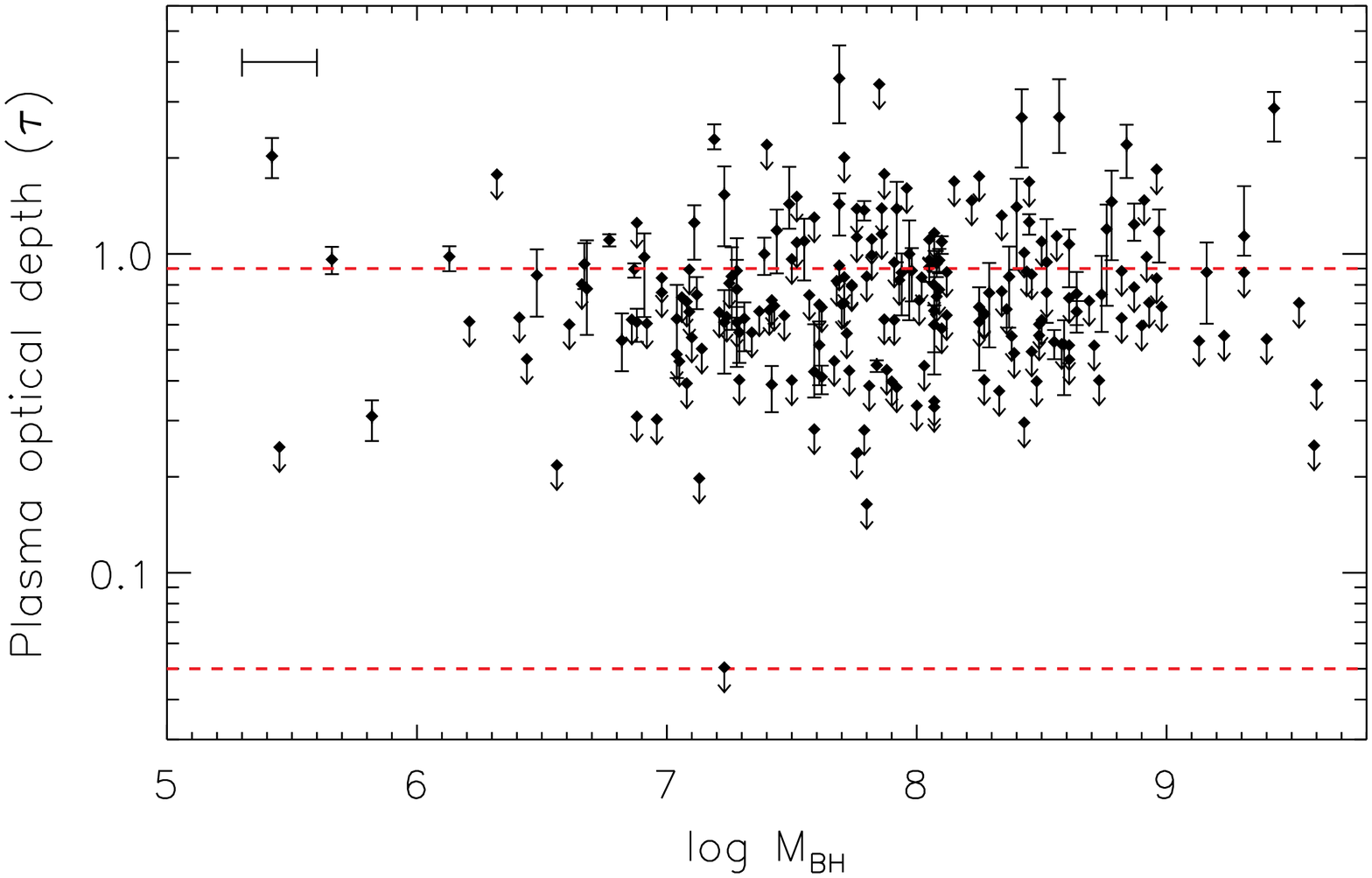}
\includegraphics[width=0.48\textwidth]{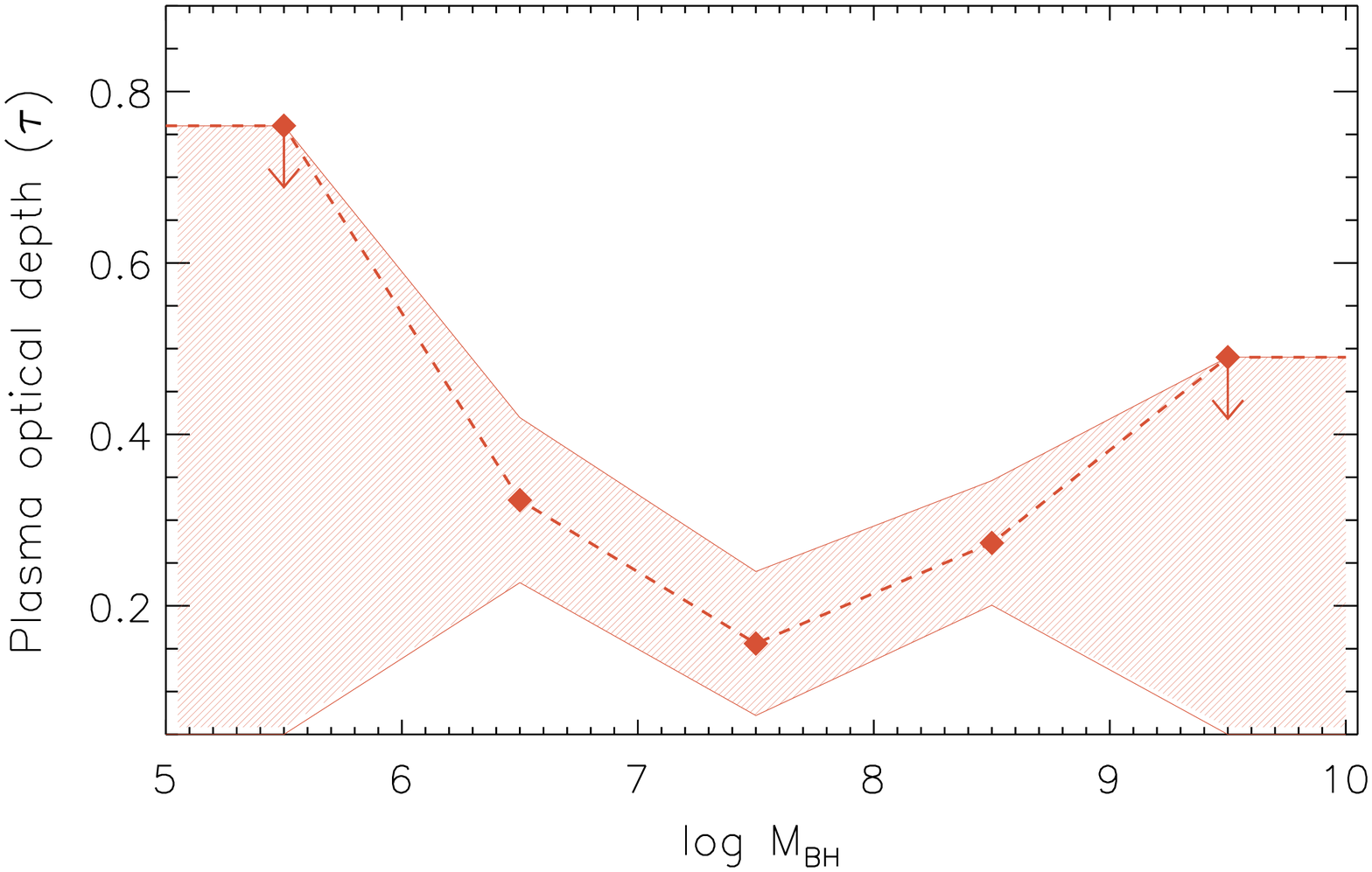}
  \caption{{\it Left panel:} Optical depth versus $M_{\rm BH}$ (in $M_{\odot}$). The red dashed lines show the interval of $\tau$ shown in the right panel. The bar in the top left corner shows the typical uncertainty of $M_{\rm BH}$. {\it Right panel:} Median values of $\tau$ for different intervals of $M_{\rm BH}$. The shaded area represents to the median absolute deviation.}
\label{fig:tau_vsMBH}
\end{figure*}

\begin{figure*}
\centering
\includegraphics[width=0.48\textwidth]{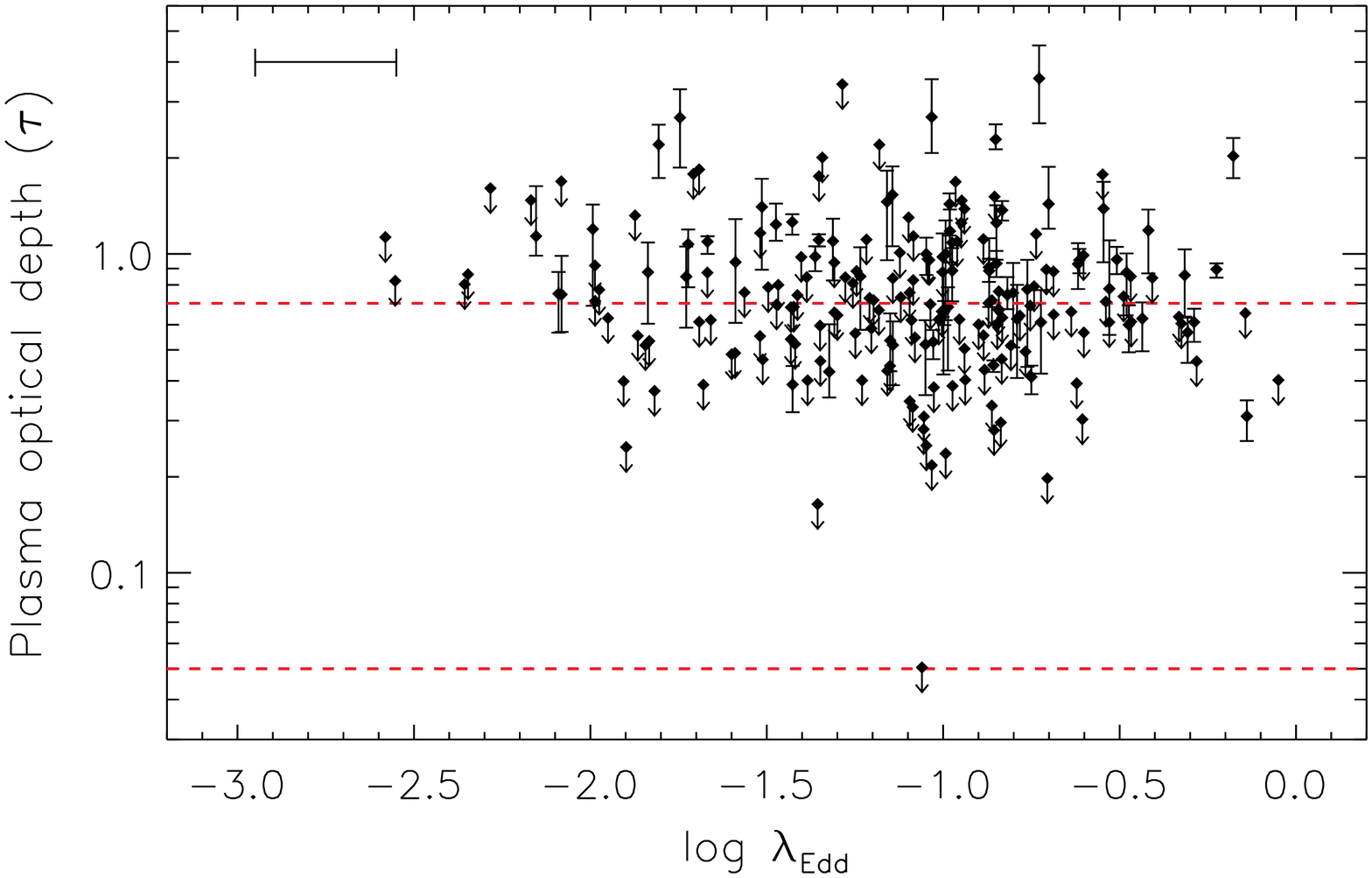}
\includegraphics[width=0.48\textwidth]{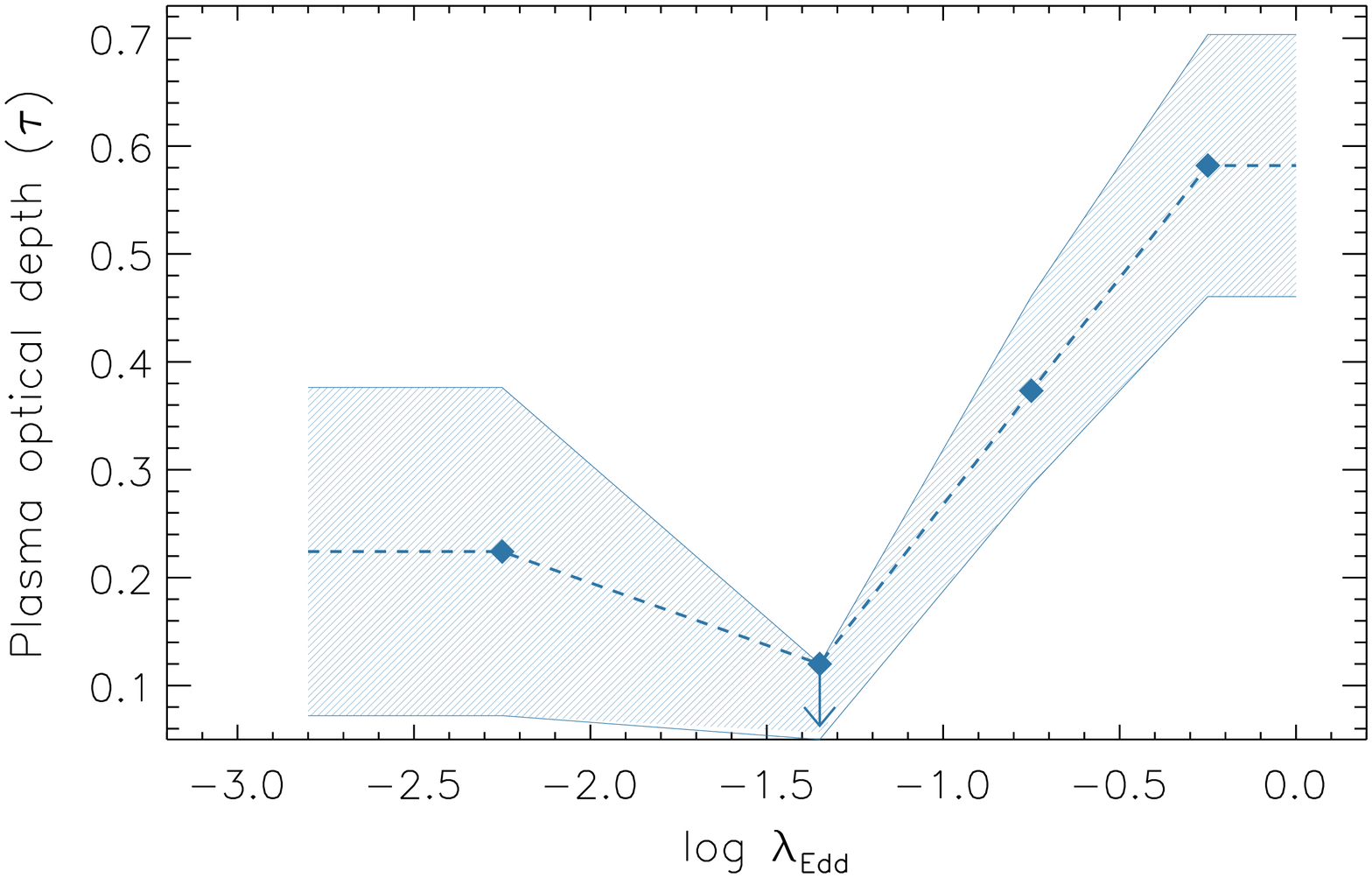}
  \caption{{\it Left panel:} Optical depth of the plasma versus the Eddington ratio. The red dashed lines show the interval of $\tau$ shown in the right panel. The bar in the top left corner shows the typical uncertainty of $\lambda_{\rm Edd}$. {\it Right panel:} Median values of $\tau$ for different intervals of $\lambda_{\rm Edd}$. The shaded area corresponds to the median absolute deviation.}
\label{fig:tau_vsEddratio}
\end{figure*}

To calculate the relation between the 2--10\,keV photon index $\Gamma$, $kT_{\rm\,e}$ and $\tau$ we simulated 10,000 spectra using the \textsc{compps} model \citep{Poutanen:1996dn} in \textsc{XSPEC} \citep{Arnaud:1996kx}, which produces X-ray spectra from Comptonization in a plasma with variable geometry, temperature and optical depth. We assumed a slab geometry, and created a uniform grid in the ranges $0.1 \leq \tau \leq 5.1$ and $30 \leq (kT_{\rm e}/\rm keV)\leq 275$. We set the inclination angle to $45^{\circ}$, and only considered the primary X-ray emission, setting the reflection parameter to $R=0$. The seed photons were produced using a multi-color disk with an inner disk temperature of $10$\,eV. The photon index was inferred, for each value of $kT_{\rm e}$ and $\tau$, by fitting the simulated spectra with a power-law with a simple powerlaw model (\textsc{pow}) in the 2--10\,keV range, leaving both the normalization and photon index free to vary. We then fit the data with:
\begin{equation}\label{eq:Gamma_kt_tau}
\Gamma=d+e\times \log(kT_{\rm e})+f\times \log(\tau).
\end{equation}
From the fit we find $d=2.160$, $e=-0.317$, and $f=-1.062$; the median of the absolute difference between the photon index and the value found with Eq.\,\ref{eq:Gamma_kt_tau} is $\lvert \Gamma-[d+e\times \log(kT_{\rm e})+f\times log(\tau)] \rvert=0.04$, showing that the fit can reproduce well the data.

We can then invert Eq.\,\ref{eq:Gamma_kt_tau} to obtain the optical depth as a function of $kT_{\rm e}$ and $\Gamma$:
\begin{equation}\label{eq:tau_gamma_kte}
\tau=10^{\frac{\Gamma-d}{f}}\times (kT_{\rm e})^{-0.3}.
\end{equation}
From our spectral analysis we have both $\Gamma$ and $kT_{\rm e}=E_{\rm C}/2$ (see \S\ref{sec:thetalplane}), so that we can calculate $\tau$ for the 211 AGN in our sample.  The sources for which only a lower limit on $E_{\rm C}$ is available have upper limits on $\tau$. To be consistent with the simulations we used the photon index obtained by fitting the $E\leq 10$\,keV spectrum (see \citealp{Ricci:2017bf} for details).
The distribution of the plasma optical depth for our sample is shown in Fig.\,\ref{fig:tau_histo}. Using \textsc{asurv}, following the same approach outlined in \S\,\ref{sec:Ecvsaccretion} we find that, for the whole sample, the median optical depth is $\tau=0.25 \pm 0.06$. 

We investigated the relation between $\tau$, the X-ray luminosity (Fig.\,\ref{fig:tau_vsL}), the black hole mass (Fig.\,\ref{fig:tau_vsMBH}) and the Eddington ratio (Fig.\,\ref{fig:tau_vsEddratio}), and found no statistically significant correlations between these quantities. However, we find a $\simeq 3\sigma$ difference in the optical depth of objects accreting at low ($\lambda_{\rm Edd} \leq 0.1$) and at high ($\lambda_{\rm Edd} > 0.1$) Eddington ratios, with the median values being $\tau =0.15\pm 0.07$ and $\tau=0.44\pm 0.07$, respectively. If the optical depth of the corona increases with the density of the accretion disk ($n$), then the increase of $\tau$ with $\lambda_{\rm Edd}$ would be consistent with the classical accretion disk model \citep{Shakura:1973if}, according to which $n\propto \lambda_{\rm\,Edd}$.
In a recent work, \cite{Tortosa:2018rm} found an anti-correlation between the temperature and the optical depth of the X-ray emitting plasma. Considering this, and the decrease of the temperature of the Comptonizing plasma with the Eddington ratio, one would naturally expect that at higher $\lambda_{\rm Edd}$ AGN tend to preferentially have coronae with larger optical depths.

\section{The temperature-compactness plane and the $\Gamma-\lambda_{\rm Edd}$ correlation}\label{sec:GammaEddratiocorrelation}

\subsection{The $\Gamma-\lambda_{\rm Edd}$ relation}
A relation between the photon index and the Eddington ratio has been reported by several authors over the past two decades (e.g., \citealp{Brandt:1997fr,Shemmer:2006bs,Shemmer:2008jy,Risaliti:2009bh,Fanali:2013wd,Brightman:2013jy,Brightman:2016qr,Kawamuro:2016uo}), which have shown that, for increasing $\lambda_{\rm Edd}$, the X-ray continuum tend to be steeper. 
Most of these works have found that the correlation 
\begin{equation}\label{eq:gammaEddratio}
\Gamma=\psi \log\lambda_{\rm Edd}+ \omega
\end{equation}
has a slope $\psi \sim 0.3$ (e.g., \citealp{Shemmer:2008jy,Brightman:2013jy}), while a steeper slope ($\psi \simeq 0.6$) was reported by \cite{Risaliti:2009bh}, who studied SDSS quasars with archival {\it XMM-Newton} observations. More recently, \cite{Trakhtenbrot:2017bh}, using BASS, found instead a significantly weaker and flatter ($\psi \simeq 0.15$) correlation when using $\Gamma$ obtained by considering complex spectral models (see \citealp{Ricci:2017bf} for details). Interestingly, when using $\Gamma$ obtained by fitting the spectra of unobscured AGN with a simple power law model in the 2--10\,keV range, \citet{Trakhtenbrot:2017bh} found a slope similar ($\phi=0.30 \pm 0.09$) to that reported by previous studies.
The existence of a relation between $\Gamma$ and $\lambda_{\rm Edd}$ has been confirmed by repeated observations of individual sources, which have shown that the photon index increases with the flux (e.g., \citealp{Perola:1986le,Matsuoka:1990tw,Lamer:2003fq,Sobolewska:2009dp}). Interestingly, \citet{Sobolewska:2009dp} found that $\psi$ differs from object to object, varying from $\simeq 0.10$ to $\simeq 0.30$, and that the slope for the average spectral slope versus the average Eddington ratio is $\psi=0.08\pm0.02$. This slope is consistent with that found for BASS by \cite{Trakhtenbrot:2017bh}, and with the value reported by \citeauthor{Ricci:2013oq} (\citeyear{Ricci:2013oq}; $\psi=0.12\pm0.04$) for a sample of 36 nearby AGN, considering the average $\Gamma$ and $\lambda_{\rm Edd}$.
The difference between the slopes found by the works reported above is likely related to the approach used for the spectral fitting (i.e., a simple power-law model or more complex models), to the energy band, and to the sample used. 

\begin{figure}
\centering
\includegraphics[width=0.48\textwidth]{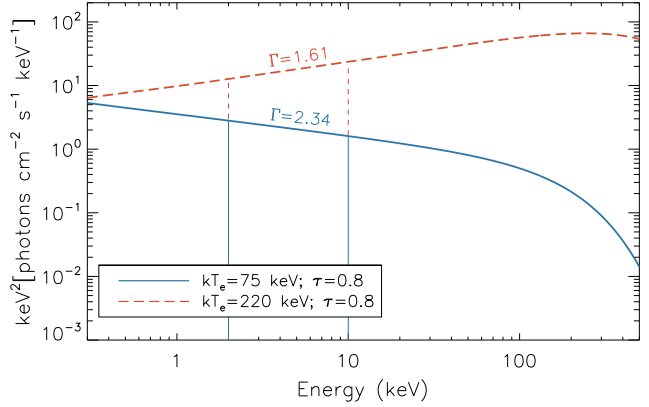}
  \caption{Comptonization X-ray spectra obtained using the \textsc{compps} model assuming an optical depth $\tau=0.8$, a spherical corona and different plasma temperatures: $kT_{\rm e}=75$\,keV (blue continuous line) and $kT_{\rm e}=220$\,keV (red dashed line). The photon indices obtained by fitting the spectra with a simple power law model in the 2--10\,keV range are also reported, showing that the X-ray continuum becomes harder for higher temperatures of the corona (see \S\ref{sec:GammaEddratiocorrelation}).}
\label{fig:compps_spec}
\end{figure}

\subsection{Explaining the $\Gamma-\lambda_{\rm Edd}$ relation with the pair line}\label{sec:simulations}

The physical mechanism responsible for this correlation is still debated. It has been proposed that this might be related to a more efficient cooling of the X-ray emitting plasma at higher $\lambda_{\rm Edd}$, due to the larger amount of optical and UV seed photons produced by the accretion disk (e.g., \citealp{Vasudevan:2007qt,Davis:2011kk}). 
However, as argued by \cite{Trakhtenbrot:2017bh}, the number of optical and UV photons also increases when the black hole mass decreases (e.g., \citealp{Done:2012pb,Slone:2012qa}), so that one would also expect a relation between $\Gamma$ and $M_{\rm BH}$, which is not observed. 
In the previous sections we have shown that the temperature of the Comptonizing plasma tends to decrease for increasing $\lambda_{\rm\,Edd}$ (\S\ref{sec:Ecvsaccretion}), and that this effect could be related to the fact that coronae tend to concentrate around the runaway pair creation line (\S\ref{sec:thetalplane}). This could provide an alternative mechanism for the $\Gamma-\lambda_{\rm Edd}$ relation, since the photon index depends on the temperature of the plasma (see Fig.\,\ref{fig:compps_spec} and Eq.\ref{eq:tau_gamma_kte}).

To test whether the limits imposed by pair production on the plasma temperature for a given compactness parameter could explain the observed relation between $\Gamma$ and $\lambda_{\rm Edd}$, we first interpolated the limit of the runaway pair production region in the $\Theta-l$ diagram (considering a slab corona, see \citealp{Stern:1995sh}) using a polynomial of the second order:
\begin{equation}\label{eq:thetalrel}
\log \Theta=a+ b\times\log l + c\times\log^2 l.
\end{equation}
From the fit we obtained a=$-0.282164$, b=$-0.239618$ and c=$0.0215106$. We then assumed that typically coronae are distributed along this line, as shown in the right panel of Fig.\,\ref{fig:Ecut_thetal}. 
We simulated 10,000 spectra using \textsc{compps}, similarly to what was done in \S\ref{sect:tau}. We set the plasma temperature $kT_{\rm e}$ to depend on the Eddington ratio by combining Eqs.\,\ref{eq:l} and \ref{eq:theta} with Eq.\,\ref{eq:thetalrel}, i.e. transforming the $\Theta(l)$ relation into a $kT(\lambda_{\rm Edd})$ function:
\begin{equation}
\log (kT_{\rm e})=a_1+ b\times\log (\eta\lambda_{\rm Edd}) + c\times\log^2 (\eta\lambda_{\rm Edd}),
\end{equation}
where $a_1=a+\log(m_{\rm e}c^2)$ and $\eta=\frac{\pi m_{\rm p} }{50m_{e}}$, assuming $R_{\rm X}=10\,R_{\rm g}$ and $\kappa_{\rm x}=20$.
We explored a range in Eddington ratio between $10^{-3}$ and 1, which translates into a plasma temperature interval of $kT_{\rm e}=220-73$\,keV. We explored a range of Comptonizing plasma optical depths ($\tau$), from 0.1 to 1.5, and two different geometries of the corona (sphere and slab). In Figure\,\ref{fig:compps_spec} we show, as an example, two spectra obtained with \textsc{compps}, assuming the parameters reported above, an optical depth of $\tau=0.8$ and plasma temperatures of $kT_{\rm e}=220$\,keV (red dashed line) and $kT_{\rm e}=75$\,keV (blue continuous line), which encompass the range of temperature explored in our simulations. The figure clearly shows that cooler plasma tend to create significantly steeper X-ray spectra in the 2--10\,keV range.

\begin{figure}
\centering
\includegraphics[width=0.48\textwidth]{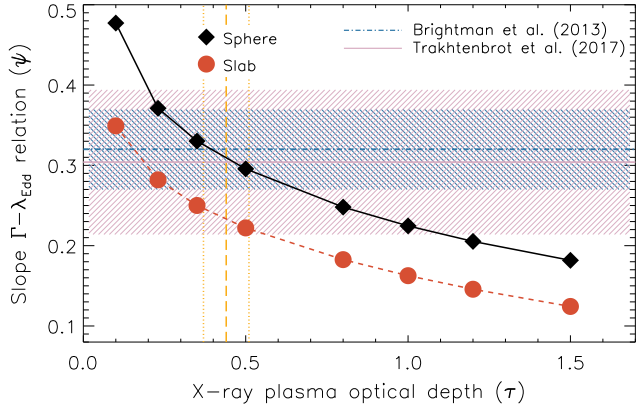}
  \caption{Slopes of the $\Gamma-\lambda_{\rm Edd}$ relation ($\psi$) obtained by simulating a population of AGN with coronae following the pair line in the $\Theta-l$ diagram, as described in \S\ref{sec:GammaEddratiocorrelation}, for different values of the plasma optical depth ($\tau$), and for two geometries of the X-ray source: sphere (black diamonds) and slab (red circles). The horizontal lines show the slopes obtained by recent works, while the shaded areas illustrate their uncertainties. The vertical orange dashed line shows the median optical depth in our sample (\S\ref{sect:tau}), while the dotted orange lines its 1$\sigma$ uncertainty. The simulations show that a temperature of the X-ray emitting plasma depending on the Eddington ratio following Eqs.\,\ref{eq:l}, \ref{eq:theta} and \ref{eq:thetalrel} can reproduce the $\Gamma-\lambda_{\rm Edd}$ correlation for a large range of optical depths.  }
\label{fig:slopeEddratioGamma}
\end{figure}

The simulated spectra were then fit with a powerlaw model in the 2--10\,keV range. We studied the relation between $\Gamma$ and $\lambda_{\rm Edd}$, fitting the data with an expression that follows Eq.\,\ref{eq:gammaEddratio}. As shown in Fig.\,\ref{fig:slopeEddratioGamma}, the fact that coronae follow the pair line could easily reproduce the observed slope of the $\Gamma-\lambda_{\rm Edd}$ correlation. In the figure we use the value of $\phi$ from \cite{Trakhtenbrot:2017bh} inferred using the photon index obtained by applying a simple power-law model to the 2--10\,keV spectrum, consistently with what was done for our simulations. The optical depth extrapolated from our sample in \S\ref{sect:tau} would correspond to slopes in the range $\psi\simeq0.26-0.30$, in agreement with the observations (orange vertical lines in Fig.\,\ref{fig:slopeEddratioGamma}).
The steeper slopes of the correlation obtained for optically thinner plasma are likely due to the stronger influence of changes in temperatures on the X-ray spectrum.  The large scatter observed in the $\Gamma-\lambda_{\rm Edd}$ correlation (e.g., \citealp{Ho:2016pi,Trakhtenbrot:2017bh}) could be ascribed to several causes, such as the intrinsic scatter in the $\Theta-l$ diagram, different optical depths of the Comptonizing region, and/or different sizes and geometries of the corona. Moreover, pair production in non-thermal plasma could create a large range of plasma temperatures \citep{Fabian:2017dq}, which would also contribute to the scatter.

\section{Summary and conclusion}\label{sect:summary}
We have studied here the relation between the coronal and accretion properties of 211 local unobscured AGN from the BAT AGN Spectroscopic Survey. The main findings of our work are the following.
\begin{itemize}

\item The median temperature of the X-ray emitting plasma for the objects in our sample is $kT_{\rm e}=105\pm18$\,keV.

\item The main parameter driving the cutoff energy is the Eddington ratio (see $\S$\ref{sec:Ecvsaccretion}, Fig.\,\ref{fig:Ecut_vsEddratio} and Fig.\,\ref{fig:Ecut_vsEddratio_othPar}). This is shown by the negative correlation between $E_{\rm C}$ and $\lambda_{\rm Edd}$ (Fig.\,\ref{fig:Ecut_vsEddratio}), and by the fact that any trend with luminosity or black hole mass disappears when dividing the samples into bins of $\lambda_{\rm Edd}$, while the difference between low and high Eddington ratio sources is always observed, regardless of the interval of luminosity (right panel of Fig.\,\ref{fig:Ecut_vsLum_othPar}) or black hole mass (right panel of Fig.\,\ref{fig:Ecut_vsMBH_othPar}). 
\item At low Eddington ratios ($\lambda_{\rm Edd}\leq 0.1$) the median cutoff energy is $E_{\rm C}=370 \pm 51$\, keV, while at high Eddington ratios ($\lambda_{\rm Edd}> 0.1$) is $E_{\rm C}=160 \pm 41$\,keV, which implies a $3.2\sigma$ difference between the two subsamples.

\item We studied the distribution of the AGN in our sample in the temperature-compactness ($\Theta-l$) parameter space ($\S$\ref{sec:thetalplane}), and found that AGN typically tend to avoid the pair runaway region, and to lie between the $e^{-}-e^{-}$ coupling line and the pair line for a slab corona (Eq.\,\ref{eq:thetalrel}), implying that the geometry of the corona may be better described as a slab (instead of a sphere).

\item The relation between $E_{\rm C}$ and $\lambda_{\rm Edd}$ can be explained by the fact that AGN tend to avoid the pair runaway region in the $\Theta-l$ diagram, considering that, for a fixed size of the X-ray emitting region, the compactness is proportional to the Eddington ratio ($l\propto \lambda_{\rm Edd}$, see Eq.\,\ref{eq:l}).

\item Using spectral simulations, considering a slab corona, we show that the optical depth of the Comptonizing plasma can be calculated from $\Gamma$ and $E_{\rm C}$ using Eq.\,\ref{eq:tau_gamma_kte} (see \S\ref{sect:tau}). The median value of the optical depth for our sample is $\tau=0.25 \pm 0.06$, and objects accreting at $\lambda_{\rm Edd} \leq 0.1$ have a lower median optical depth ($\tau =0.15\pm 0.07$) than those with $\lambda_{\rm Edd} > 0.1$ ($\tau=0.44\pm 0.07$).

\item Simulating AGN populations with an X-ray spectral Comptonization model, we showed that Comptonizing plasma with temperatures and compactness lying along the pair line can straightforwardly reproduce the observed slope of the $\Gamma-\lambda_{\rm Edd}$ relation (see \S\ref{sec:GammaEddratiocorrelation}).

\end{itemize}

BASS aims to reliably estimate, in the near future, black hole masses for about 1,000 local AGN. Therefore future studies of a larger number of hard X-ray selected AGN carried out with {\it Swift}/BAT and {\it NuSTAR} (in the framework of the BAT legacy survey), will be able to better characterise the relation between the cutoff energy and the Eddington ratio, and to understand the importance of non-thermal components in the X-ray emitting plasma.

\section*{Acknowledgements}
We thank the reviewer for the useful comments, which helped us improve the quality of the manuscript.
This work made use of data from the NASA/IPAC Infrared Science Archive and NASA/IPAC Extragalactic Database (NED), which are operated by the Jet Propulsion Laboratory, California Institute of Technology, under contract with the National Aeronautics and Space Administration. We acknowledge financial support from the National Key R\&D Program of China grant No. 2016YFA0400702 (LH), the National Science Foundation of China grants No. 11473002 and 1721303 (LH), FONDECYT 1141218 (CR, FEB), FONDECYT 1160999 (ET), CONICYT PIA ACT172033 (ET), Basal-CATA PFB--06/2007 (CR, FEB, ET), the China-CONICYT fund (CR), the CONICYT+PAI Convocatoria Nacional subvencion a instalacion en la academia convocatoria a\~{n}o 2017 PAI77170080 (CR), the Swiss National Science Foundation (Grant PP00P2\_138979 and PP00P2\_166159, KS), the Swiss National Science Foundation (SNSF) through the Ambizione fellowship grant PZ00P2\textunderscore154799/1 (MK), the NASA ADAP award NNH16CT03C (MK), the ERC Advanced Grant Feedback 340442 (ACF), and the Ministry of Economy, Development, and Tourism's Millennium Science Initiative through grant IC120009, awarded to The Millennium Institute of Astrophysics, MAS (FEB). This work was partly supported by the Grant-in-Aid for Scientific Research 17K05384 (YU) from the Ministry of Education, Culture, Sports, Science and Technology of Japan (MEXT).

\bibliographystyle{mnras}
 \bibliography{cutoff_bass}

\begin{thebibliography}{117}
\expandafter\ifx\csname natexlab\endcsname\relax\def\natexlab#1{#1}\fi

\bibitem[{Arnaud}(1996)]{Arnaud:1996kx}
{Arnaud} K.~A., 1996, in { Astronomical Data Analysis Software and Systems
  V\/}, edited by {G.~H.~Jacoby \& J.~Barnes}, vol. 101 of { Astronomical
  Society of the Pacific Conference Series\/}, ~17

\bibitem[{Ballantyne}(2014)]{Ballantyne:2014dq}
{Ballantyne} D.~R., 2014, \mnras, 437, 2845

\bibitem[{Ballantyne} et~al.(2014){Ballantyne}, {Bollenbacher}, {Brenneman}
  et~al.]{Ballantyne:2014kc}
{Ballantyne} D.~R., {Bollenbacher} J.~M., {Brenneman} L.~W., et~al., 2014,
  \apj, 794, 62

\bibitem[{Balokovi{\'c}} et~al.(2015){Balokovi{\'c}}, {Matt}, {Harrison}
  et~al.]{Balokovic:2015mi}
{Balokovi{\'c}} M., {Matt} G., {Harrison} F.~A., et~al., 2015, \apj, 800, 62

\bibitem[{Barthelmy} et~al.(2005){Barthelmy}, {Barbier}, {Cummings}
  et~al.]{Barthelmy:2005uq}
{Barthelmy} S.~D., {Barbier} L.~M., {Cummings} J.~R., et~al., 2005, \ssr, 120,
  143

\bibitem[{Baumgartner} et~al.(2013){Baumgartner}, {Tueller}, {Markwardt}
  et~al.]{Baumgartner:2013ee}
{Baumgartner} W.~H., {Tueller} J., {Markwardt} C.~B., et~al., 2013, \apjs, 207,
  19

\bibitem[{Beckmann} et~al.(2009){Beckmann}, {Soldi}, {Ricci}
  et~al.]{Beckmann:2009fk}
{Beckmann} V., {Soldi} S., {Ricci} C., et~al., 2009, \aap, 505, 417

\bibitem[{Berney} et~al.(2015){Berney}, {Koss}, {Trakhtenbrot}
  et~al.]{Berney:2015uq}
{Berney} S., {Koss} M., {Trakhtenbrot} B., et~al., 2015, \mnras, 454, 3622

\bibitem[{Bisnovatyi-Kogan} et~al.(1971){Bisnovatyi-Kogan}, {Zel'dovich} \&
  {Syunyaev}]{Bisnovatyi-Kogan:1971nr}
{Bisnovatyi-Kogan} G.~S., {Zel'dovich} Y.~B., {Syunyaev} R.~A., 1971, \sovast,
  15, 17

\bibitem[{Brandt} et~al.(1997){Brandt}, {Mathur} \& {Elvis}]{Brandt:1997fr}
{Brandt} W.~N., {Mathur} S., {Elvis} M., 1997, \mnras, 285, L25

\bibitem[{Brenneman} et~al.(2014){Brenneman}, {Madejski}, {Fuerst}
  et~al.]{Brenneman:2014mi}
{Brenneman} L.~W., {Madejski} G., {Fuerst} F., et~al., 2014, \apj, 788, 61

\bibitem[{Brightman} et~al.(2016){Brightman}, {Masini}, {Ballantyne}
  et~al.]{Brightman:2016qr}
{Brightman} M., {Masini} A., {Ballantyne} D.~R., et~al., 2016, \apj, 826, 93

\bibitem[{Brightman} et~al.(2013){Brightman}, {Silverman}, {Mainieri}
  et~al.]{Brightman:2013jy}
{Brightman} M., {Silverman} J.~D., {Mainieri} V., et~al., 2013, \mnras, 433,
  2485

\bibitem[{Cavaliere} \& {Morrison}(1980)]{Cavaliere:1980cs}
{Cavaliere} A., {Morrison} P., 1980, \apjl, 238, L63

\bibitem[{Chartas} et~al.(2009){Chartas}, {Kochanek}, {Dai}, {Poindexter} \&
  {Garmire}]{Chartas:2009sy}
{Chartas} G., {Kochanek} C.~S., {Dai} X., {Poindexter} S., {Garmire} G., 2009,
  \apj, 693, 174

\bibitem[{Dadina}(2007)]{Dadina:2007sj}
{Dadina} M., 2007, \aap, 461, 1209

\bibitem[{Davis} \& {Laor}(2011)]{Davis:2011kk}
{Davis} S.~W., {Laor} A., 2011, \apj, 728, 98

\bibitem[{De Marco} et~al.(2013){De Marco}, {Ponti}, {Cappi}
  et~al.]{De-Marco:2013fx}
{De Marco} B., {Ponti} G., {Cappi} M., et~al., 2013, \mnras, 431, 2441

\bibitem[{de Rosa} et~al.(2012){de Rosa}, {Panessa}, {Bassani}
  et~al.]{de-Rosa:2012pd}
{de Rosa} A., {Panessa} F., {Bassani} L., et~al., 2012, \mnras, 420, 2087

\bibitem[{Done} et~al.(2012){Done}, {Davis}, {Jin}, {Blaes} \&
  {Ward}]{Done:2012pb}
{Done} C., {Davis} S.~W., {Jin} C., {Blaes} O., {Ward} M., 2012, \mnras, 420,
  1848

\bibitem[{Fabian}(1994)]{Fabian:1994ht}
{Fabian} A.~C., 1994, \apjs, 92, 555

\bibitem[{Fabian}(2012)]{Fabian:2012eq}
{Fabian} A.~C., 2012, \araa, 50, 455

\bibitem[{Fabian} et~al.(2017){Fabian}, {Lohfink}, {Belmont}, {Malzac} \&
  {Coppi}]{Fabian:2017dq}
{Fabian} A.~C., {Lohfink} A., {Belmont} R., {Malzac} J., {Coppi} P., 2017,
  \mnras, 467, 2566

\bibitem[{Fabian} et~al.(2015){Fabian}, {Lohfink}, {Kara}, {Parker},
  {Vasudevan} \& {Reynolds}]{Fabian:2015db}
{Fabian} A.~C., {Lohfink} A., {Kara} E., {Parker} M.~L., {Vasudevan} R.,
  {Reynolds} C.~S., 2015, \mnras, 451, 4375

\bibitem[{Fabian} et~al.(2009){Fabian}, {Zoghbi}, {Ross} et~al.]{Fabian:2009hi}
{Fabian} A.~C., {Zoghbi} A., {Ross} R.~R., et~al., 2009, \nat, 459, 540

\bibitem[{Fanali} et~al.(2013){Fanali}, {Caccianiga}, {Severgnini}
  et~al.]{Fanali:2013wd}
{Fanali} R., {Caccianiga} A., {Severgnini} P., et~al., 2013, \mnras, 433, 648

\bibitem[{Feigelson} \& {Nelson}(1985)]{Feigelson:1985qv}
{Feigelson} E.~D., {Nelson} P.~I., 1985, \apj, 293, 192

\bibitem[{Ferrarese} \& {Merritt}(2000)]{Ferrarese:2000kq}
{Ferrarese} L., {Merritt} D., 2000, \apjl, 539, L9

\bibitem[{Gebhardt} et~al.(2000){Gebhardt}, {Bender}, {Bower}
  et~al.]{Gebhardt:2000fj}
{Gebhardt} K., {Bender} R., {Bower} G., et~al., 2000, \apjl, 539, L13

\bibitem[{Gehrels} et~al.(2004){Gehrels}, {Chincarini}, {Giommi}
  et~al.]{Gehrels:2004kx}
{Gehrels} N., {Chincarini} G., {Giommi} P., et~al., 2004, \apj, 611, 1005

\bibitem[{Ghisellini} et~al.(1993){Ghisellini}, {Haardt} \&
  {Fabian}]{Ghisellini:1993fj}
{Ghisellini} G., {Haardt} F., {Fabian} A.~C., 1993, \mnras, 263, L9

\bibitem[{Gilli} et~al.(2007){Gilli}, {Comastri} \& {Hasinger}]{Gilli:2007yg}
{Gilli} R., {Comastri} A., {Hasinger} G., 2007, \aap, 463, 79

\bibitem[{Greene} \& {Ho}(2005)]{Greene:2005wf}
{Greene} J.~E., {Ho} L.~C., 2005, \apj, 630, 122

\bibitem[{Guilbert} et~al.(1983){Guilbert}, {Fabian} \&
  {Rees}]{Guilbert:1983ek}
{Guilbert} P.~W., {Fabian} A.~C., {Rees} M.~J., 1983, \mnras, 205, 593

\bibitem[{Haardt} \& {Maraschi}(1991)]{Haardt:1991qr}
{Haardt} F., {Maraschi} L., 1991, \apjl, 380, L51

\bibitem[{Haardt} \& {Maraschi}(1993)]{Haardt:1993cv}
{Haardt} F., {Maraschi} L., 1993, \apj, 413, 507

\bibitem[{Haardt} et~al.(1994){Haardt}, {Maraschi} \&
  {Ghisellini}]{Haardt:1994ys}
{Haardt} F., {Maraschi} L., {Ghisellini} G., 1994, \apjl, 432, L95

\bibitem[{Harrison} et~al.(2013){Harrison}, {Craig}, {Christensen}
  et~al.]{Harrison:2013lq}
{Harrison} F.~A., {Craig} W.~W., {Christensen} F.~E., et~al., 2013, \apj, 770,
  103

\bibitem[{Ho} \& {Kim}(2016)]{Ho:2016pi}
{Ho} L.~C., {Kim} M., 2016, \apj, 821, 48

\bibitem[{Isobe} et~al.(1986){Isobe}, {Feigelson} \& {Nelson}]{Isobe:1986ys}
{Isobe} T., {Feigelson} E.~D., {Nelson} P.~I., 1986, \apj, 306, 490

\bibitem[{Johnson} et~al.(1997){Johnson}, {McNaron-Brown}, {Kurfess},
  {Zdziarski}, {Magdziarz} \& {Gehrels}]{Johnson:1997hw}
{Johnson} W.~N., {McNaron-Brown} K., {Kurfess} J.~D., {Zdziarski} A.~A.,
  {Magdziarz} P., {Gehrels} N., 1997, \apj, 482, 173

\bibitem[{Kara} et~al.(2013){Kara}, {Fabian}, {Cackett}, {Uttley}, {Wilkins} \&
  {Zoghbi}]{Kara:2013wu}
{Kara} E., {Fabian} A.~C., {Cackett} E.~M., {Uttley} P., {Wilkins} D.~R.,
  {Zoghbi} A., 2013, \mnras, 434, 1129

\bibitem[{Kara} et~al.(2017){Kara}, {Garc{\'{\i}}a}, {Lohfink}
  et~al.]{Kara:2017lq}
{Kara} E., {Garc{\'{\i}}a} J.~A., {Lohfink} A., et~al., 2017, \mnras, 468, 3489

\bibitem[{Kawamuro} et~al.(2016){Kawamuro}, {Ueda}, {Tazaki}, {Ricci} \&
  {Terashima}]{Kawamuro:2016uo}
{Kawamuro} T., {Ueda} Y., {Tazaki} F., {Ricci} C., {Terashima} Y., 2016, \apjs,
  225, 14

\bibitem[{King} \& {Pounds}(2015)]{King:2015ys}
{King} A., {Pounds} K., 2015, \araa, 53, 115

\bibitem[{Kormendy} \& {Ho}(2013)]{Kormendy:2013uf}
{Kormendy} J., {Ho} L.~C., 2013, \araa, 51, 511

\bibitem[{Koss} et~al.(2017){Koss}, {Trakhtenbrot}, {Ricci}
  et~al.]{Koss:2017fp}
{Koss} M., {Trakhtenbrot} B., {Ricci} C., et~al., 2017, \apj, 850, 74

\bibitem[{Lamer} et~al.(2003){Lamer}, {McHardy}, {Uttley} \&
  {Jahoda}]{Lamer:2003fq}
{Lamer} G., {McHardy} I.~M., {Uttley} P., {Jahoda} K., 2003, \mnras, 338, 323

\bibitem[{Lamperti} et~al.(2017){Lamperti}, {Koss}, {Trakhtenbrot}
  et~al.]{Lamperti:2017kq}
{Lamperti} I., {Koss} M., {Trakhtenbrot} B., et~al., 2017, \mnras

\bibitem[{Lanzuisi} et~al.(2016){Lanzuisi}, {Perna}, {Comastri}
  et~al.]{Lanzuisi:2016pl}
{Lanzuisi} G., {Perna} M., {Comastri} A., et~al., 2016, \aap, 590, A77

\bibitem[{Liu} et~al.(2015){Liu}, {Taam}, {Qiao} \& {Yuan}]{Liu:2015vn}
{Liu} B.~F., {Taam} R.~E., {Qiao} E., {Yuan} W., 2015, \apj, 806, 223

\bibitem[{Liu} et~al.(2017){Liu}, {Taam}, {Qiao} \& {Yuan}]{Liu:2017qy}
{Liu} B.~F., {Taam} R.~E., {Qiao} E., {Yuan} W., 2017, \apj, 847, 96

\bibitem[{Lohfink} et~al.(2017){Lohfink}, {Fabian}, {Ballantyne}
  et~al.]{Lohfink:2017bq}
{Lohfink} A.~M., {Fabian} A.~C., {Ballantyne} D.~R., et~al., 2017, \apj, 841,
  80

\bibitem[{Lohfink} et~al.(2015){Lohfink}, {Ogle}, {Tombesi}
  et~al.]{Lohfink:2015ec}
{Lohfink} A.~M., {Ogle} P., {Tombesi} F., et~al., 2015, \apj, 814, 24

\bibitem[{Lubi{\'n}ski} et~al.(2016){Lubi{\'n}ski}, {Beckmann}, {Gibaud}
  et~al.]{Lubinski:2016ao}
{Lubi{\'n}ski} P., {Beckmann} V., {Gibaud} L., et~al., 2016, \mnras, 458, 2454

\bibitem[{Lubi{\'n}ski} et~al.(2010){Lubi{\'n}ski}, {Zdziarski}, {Walter}
  et~al.]{Lubinski:2010rb}
{Lubi{\'n}ski} P., {Zdziarski} A.~A., {Walter} R., et~al., 2010, \mnras, 408,
  1851

\bibitem[{Lusso} et~al.(2012){Lusso}, {Comastri}, {Simmons}
  et~al.]{Lusso:2012it}
{Lusso} E., {Comastri} A., {Simmons} B.~D., et~al., 2012, \mnras, 425, 623

\bibitem[{Malizia} et~al.(2014){Malizia}, {Molina}, {Bassani}
  et~al.]{Malizia:2014zt}
{Malizia} A., {Molina} M., {Bassani} L., et~al., 2014, \apjl, 782, L25

\bibitem[{Marinucci} et~al.(2014){Marinucci}, {Matt}, {Miniutti}
  et~al.]{Marinucci:2014fu}
{Marinucci} A., {Matt} G., {Miniutti} G., et~al., 2014, \apj, 787, 83

\bibitem[{Matsuoka} et~al.(1990){Matsuoka}, {Piro}, {Yamauchi} \&
  {Murakami}]{Matsuoka:1990tw}
{Matsuoka} M., {Piro} L., {Yamauchi} M., {Murakami} T., 1990, \apj, 361, 440

\bibitem[{Matt} et~al.(2015){Matt}, {Balokovi{\'c}}, {Marinucci}
  et~al.]{Matt:2015fe}
{Matt} G., {Balokovi{\'c}} M., {Marinucci} A., et~al., 2015, \mnras, 447, 3029

\bibitem[{Matt} et~al.(2014){Matt}, {Marinucci}, {Guainazzi}
  et~al.]{Matt:2014fv}
{Matt} G., {Marinucci} A., {Guainazzi} M., et~al., 2014, \mnras, 439, 3016

\bibitem[{McHardy} et~al.(2005){McHardy}, {Gunn}, {Uttley} \&
  {Goad}]{McHardy:2005kc}
{McHardy} I.~M., {Gunn} K.~F., {Uttley} P., {Goad} M.~R., 2005, \mnras, 359,
  1469

\bibitem[{Merloni}(2003)]{Merloni:2003kx}
{Merloni} A., 2003, \mnras, 341, 1051

\bibitem[{Merloni} \& {Fabian}(2001)]{Merloni:2001qy}
{Merloni} A., {Fabian} A.~C., 2001, \mnras, 321, 549

\bibitem[{Molina} et~al.(2009){Molina}, {Bassani}, {Malizia}
  et~al.]{Molina:2009vz}
{Molina} M., {Bassani} L., {Malizia} A., et~al., 2009, \mnras, 399, 1293

\bibitem[{Mushotzky}(1982)]{Mushotzky:1982rp}
{Mushotzky} R.~F., 1982, \apj, 256, 92

\bibitem[{Mushotzky} et~al.(1993){Mushotzky}, {Done} \&
  {Pounds}]{Mushotzky:1993bf}
{Mushotzky} R.~F., {Done} C., {Pounds} K.~A., 1993, \araa, 31, 717

\bibitem[{Nicastro} et~al.(2000){Nicastro}, {Piro}, {De Rosa}
  et~al.]{Nicastro:2000il}
{Nicastro} F., {Piro} L., {De Rosa} A., et~al., 2000, \apj, 536, 718

\bibitem[{Oh} et~al.(2017){Oh}, {Schawinski}, {Koss} et~al.]{Oh:2017zl}
{Oh} K., {Schawinski} K., {Koss} M., et~al., 2017, \mnras, 464, 1466

\bibitem[{Panessa} et~al.(2011){Panessa}, {de Rosa}, {Bassani}
  et~al.]{Panessa:2011pv}
{Panessa} F., {de Rosa} A., {Bassani} L., et~al., 2011, \mnras, 417, 2426

\bibitem[{Parker} et~al.(2014){Parker}, {Wilkins}, {Fabian}
  et~al.]{Parker:2014zp}
{Parker} M.~L., {Wilkins} D.~R., {Fabian} A.~C., et~al., 2014, \mnras, 443,
  1723

\bibitem[{Perola} et~al.(1986){Perola}, {Piro}, {Altamore}
  et~al.]{Perola:1986le}
{Perola} G.~C., {Piro} L., {Altamore} A., et~al., 1986, \apj, 306, 508

\bibitem[{Petrucci} et~al.(2000){Petrucci}, {Haardt}, {Maraschi}
  et~al.]{Petrucci:2000kq}
{Petrucci} P.~O., {Haardt} F., {Maraschi} L., et~al., 2000, \apj, 540, 131

\bibitem[{Petrucci} et~al.(2001){Petrucci}, {Haardt}, {Maraschi}
  et~al.]{Petrucci:2001nq}
{Petrucci} P.~O., {Haardt} F., {Maraschi} L., et~al., 2001, \apj, 556, 716

\bibitem[{Poutanen} \& {Svensson}(1996)]{Poutanen:1996dn}
{Poutanen} J., {Svensson} R., 1996, \apj, 470, 249

\bibitem[{Reis} \& {Miller}(2013)]{Reis:2013kq}
{Reis} R.~C., {Miller} J.~M., 2013, \apjl, 769, L7

\bibitem[{Ricci} et~al.(2013){Ricci}, {Paltani}, {Ueda} \&
  {Awaki}]{Ricci:2013oq}
{Ricci} C., {Paltani} S., {Ueda} Y., {Awaki} H., 2013, \mnras, 435, 1840

\bibitem[{Ricci} et~al.(2017{\natexlab{a}}){Ricci}, {Trakhtenbrot}, {Koss}
  et~al.]{Ricci:2017bf}
{Ricci} C., {Trakhtenbrot} B., {Koss} M.~J., et~al., 2017{\natexlab{a}}, \apjs,
  233, 17

\bibitem[{Ricci} et~al.(2017{\natexlab{b}}){Ricci}, {Trakhtenbrot}, {Koss}
  et~al.]{Ricci:2017kl}
{Ricci} C., {Trakhtenbrot} B., {Koss} M.~J., et~al., 2017{\natexlab{b}}, \nat,
  549, 488

\bibitem[{Ricci} et~al.(2015){Ricci}, {Ueda}, {Koss}, {Trakhtenbrot}, {Bauer}
  \& {Gandhi}]{Ricci:2015tg}
{Ricci} C., {Ueda} Y., {Koss} M.~J., {Trakhtenbrot} B., {Bauer} F.~E., {Gandhi}
  P., 2015, \apjl, 815, L13

\bibitem[{Ricci} et~al.(2011){Ricci}, {Walter}, {Courvoisier} \&
  {Paltani}]{Ricci:2011yw}
{Ricci} C., {Walter} R., {Courvoisier} T.~J.-L., {Paltani} S., 2011, \aap, 532,
  A102

\bibitem[{Risaliti} et~al.(2005){Risaliti}, {Elvis}, {Fabbiano}, {Baldi} \&
  {Zezas}]{Risaliti:2005kl}
{Risaliti} G., {Elvis} M., {Fabbiano} G., {Baldi} A., {Zezas} A., 2005, \apjl,
  623, L93

\bibitem[{Risaliti} et~al.(2011){Risaliti}, {Nardini}, {Salvati}
  et~al.]{Risaliti:2011yo}
{Risaliti} G., {Nardini} E., {Salvati} M., et~al., 2011, \mnras, 410, 1027

\bibitem[{Risaliti} et~al.(2009){Risaliti}, {Young} \&
  {Elvis}]{Risaliti:2009bh}
{Risaliti} G., {Young} M., {Elvis} M., 2009, \apjl, 700, L6

\bibitem[{Rybicki} \& {Lightman}(1979)]{Rybicki:1979wd}
{Rybicki} G.~B., {Lightman} A.~P., 1979, {Radiative processes in astrophysics}

\bibitem[{Schawinski} et~al.(2006){Schawinski}, {Khochfar}, {Kaviraj}
  et~al.]{Schawinski:2006kq}
{Schawinski} K., {Khochfar} S., {Kaviraj} S., et~al., 2006, \nat, 442, 888

\bibitem[{Shakura} \& {Sunyaev}(1973)]{Shakura:1973if}
{Shakura} N.~I., {Sunyaev} R.~A., 1973, \aap, 24, 337

\bibitem[{Shemmer} et~al.(2006){Shemmer}, {Brandt}, {Netzer}, {Maiolino} \&
  {Kaspi}]{Shemmer:2006bs}
{Shemmer} O., {Brandt} W.~N., {Netzer} H., {Maiolino} R., {Kaspi} S., 2006,
  \apjl, 646, L29

\bibitem[{Shemmer} et~al.(2008){Shemmer}, {Brandt}, {Netzer}, {Maiolino} \&
  {Kaspi}]{Shemmer:2008jy}
{Shemmer} O., {Brandt} W.~N., {Netzer} H., {Maiolino} R., {Kaspi} S., 2008,
  \apj, 682, 81

\bibitem[{Shimizu} et~al.(2018){Shimizu}, {Davies}, {Koss}
  et~al.]{Shimizu:2018fj}
{Shimizu} T.~T., {Davies} R.~I., {Koss} M., et~al., 2018, \apj, 856, 154

\bibitem[{Shimizu} et~al.(2016){Shimizu}, {Mel{\'e}ndez}, {Mushotzky}, {Koss},
  {Barger} \& {Cowie}]{Shimizu:2016hc}
{Shimizu} T.~T., {Mel{\'e}ndez} M., {Mushotzky} R.~F., {Koss} M.~J., {Barger}
  A.~J., {Cowie} L.~L., 2016, \mnras, 456, 3335

\bibitem[{Slone} \& {Netzer}(2012)]{Slone:2012qa}
{Slone} O., {Netzer} H., 2012, \mnras, 426, 656

\bibitem[{Sobolewska} \& {Papadakis}(2009)]{Sobolewska:2009dp}
{Sobolewska} M.~A., {Papadakis} I.~E., 2009, \mnras, 399, 1597

\bibitem[{Stern} et~al.(1995){Stern}, {Poutanen}, {Svensson}, {Sikora} \&
  {Begelman}]{Stern:1995sh}
{Stern} B.~E., {Poutanen} J., {Svensson} R., {Sikora} M., {Begelman} M.~C.,
  1995, \apjl, 449, L13

\bibitem[{Svensson}(1982{\natexlab{a}})]{Svensson:1982rw}
{Svensson} R., 1982{\natexlab{a}}, \apj, 258, 335

\bibitem[{Svensson}(1982{\natexlab{b}})]{Svensson:1982eu}
{Svensson} R., 1982{\natexlab{b}}, \apj, 258, 321

\bibitem[{Svensson}(1984)]{Svensson:1984gs}
{Svensson} R., 1984, \mnras, 209, 175

\bibitem[{Tazaki} et~al.(2011){Tazaki}, {Ueda}, {Terashima} \&
  {Mushotzky}]{Tazaki:2011xi}
{Tazaki} F., {Ueda} Y., {Terashima} Y., {Mushotzky} R.~F., 2011, \apj, 738, 70

\bibitem[{Tortosa} et~al.(2018{\natexlab{a}}){Tortosa}, {Bianchi}, {Marinucci}
  et~al.]{Tortosa:2018lk}
{Tortosa} A., {Bianchi} S., {Marinucci} A., et~al., 2018{\natexlab{a}}, \mnras,
  473, 3104

\bibitem[{Tortosa} et~al.(2018{\natexlab{b}}){Tortosa}, {Bianchi}, {Marinucci},
  {Matt} \& {Petrucci}]{Tortosa:2018rm}
{Tortosa} A., {Bianchi} S., {Marinucci} A., {Matt} G., {Petrucci} P.~O.,
  2018{\natexlab{b}}, ArXiv e-prints

\bibitem[{Tortosa} et~al.(2017){Tortosa}, {Marinucci}, {Matt}
  et~al.]{Tortosa:2017kq}
{Tortosa} A., {Marinucci} A., {Matt} G., et~al., 2017, \mnras, 466, 4193

\bibitem[{Trakhtenbrot} \& {Netzer}(2012)]{Trakhtenbrot:2012hq}
{Trakhtenbrot} B., {Netzer} H., 2012, \mnras, 427, 3081

\bibitem[{Trakhtenbrot} et~al.(2017){Trakhtenbrot}, {Ricci}, {Koss}
  et~al.]{Trakhtenbrot:2017bh}
{Trakhtenbrot} B., {Ricci} C., {Koss} M.~J., et~al., 2017, \mnras, 470, 800

\bibitem[{Treister} \& {Urry}(2005)]{Treister:2005zr}
{Treister} E., {Urry} C.~M., 2005, \apj, 630, 115

\bibitem[{Treister} et~al.(2009){Treister}, {Urry} \&
  {Virani}]{Treister:2009qa}
{Treister} E., {Urry} C.~M., {Virani} S., 2009, \apj, 696, 110

\bibitem[{Ueda} et~al.(2014){Ueda}, {Akiyama}, {Hasinger}, {Miyaji} \&
  {Watson}]{Ueda:2014ix}
{Ueda} Y., {Akiyama} M., {Hasinger} G., {Miyaji} T., {Watson} M.~G., 2014,
  \apj, 786, 104

\bibitem[{Ursini} et~al.(2015){Ursini}, {Marinucci}, {Matt}
  et~al.]{Ursini:2015dk}
{Ursini} F., {Marinucci} A., {Matt} G., et~al., 2015, \mnras, 452, 3266

\bibitem[{Vasudevan} et~al.(2013){Vasudevan}, {Brandt}, {Mushotzky}
  et~al.]{Vasudevan:2013wb}
{Vasudevan} R.~V., {Brandt} W.~N., {Mushotzky} R.~F., et~al., 2013, \apj, 763,
  111

\bibitem[{Vasudevan} \& {Fabian}(2007)]{Vasudevan:2007qt}
{Vasudevan} R.~V., {Fabian} A.~C., 2007, \mnras, 381, 1235

\bibitem[{Vasudevan} \& {Fabian}(2009)]{Vasudevan:2009ng}
{Vasudevan} R.~V., {Fabian} A.~C., 2009, \mnras, 392, 1124

\bibitem[{Winter} et~al.(2009){Winter}, {Mushotzky}, {Reynolds} \&
  {Tueller}]{Winter:2009xi}
{Winter} L.~M., {Mushotzky} R.~F., {Reynolds} C.~S., {Tueller} J., 2009, \apj,
  690, 1322

\bibitem[{Xie} et~al.(2017){Xie}, {Yuan} \& {Ho}]{Xie:2017fb}
{Xie} F.-G., {Yuan} F., {Ho} L.~C., 2017, \apj, 844, 42

\bibitem[{Xu} et~al.(2017){Xu}, {Balokovi{\'c}}, {Walton}, {Harrison},
  {Garc{\'{\i}}a} \& {Koss}]{Xu:2017kq}
{Xu} Y., {Balokovi{\'c}} M., {Walton} D.~J., {Harrison} F.~A., {Garc{\'{\i}}a}
  J.~A., {Koss} M.~J., 2017, \apj, 837, 21

\bibitem[{Zdziarski}(1985)]{Zdziarski:1985tg}
{Zdziarski} A.~A., 1985, \apj, 289, 514

\bibitem[{Zdziarski} et~al.(1996){Zdziarski}, {Johnson} \&
  {Magdziarz}]{Zdziarski:1996qp}
{Zdziarski} A.~A., {Johnson} W.~N., {Magdziarz} P., 1996, \mnras, 283, 193

\bibitem[{Zoghbi} et~al.(2012){Zoghbi}, {Fabian}, {Reynolds} \&
  {Cackett}]{Zoghbi:2012jk}
{Zoghbi} A., {Fabian} A.~C., {Reynolds} C.~S., {Cackett} E.~M., 2012, \mnras,
  422, 129

\end{thebibliography}
 
 \end{document}